\documentclass[aps,nolongbibliography,reprint]{revtex4-2}
\pdfoutput=1
\usepackage{amsmath,mathtools}
\usepackage{amssymb}
\usepackage{amsthm}
\usepackage{graphicx}
\usepackage{physics}
\usepackage{xcolor}

\def\bea#1\eea{\begin{align}#1\end{align}}
\newcommand\beq{\begin{equation}}
\newcommand\eeq{\end{equation}}
\newcommand{\e}{\varepsilon}

\begin{document}
	
\title{Nonanalytic Corrections to the Landau Diamagnetic Susceptibility In a 2D Fermi Liquid}

\author{R. David Mayrhofer}
\affiliation{School of Physics and Astronomy and William I. Fine Theoretical Physics Institute, University of Minnesota, Minneapolis, MN 55455, USA}

\author{Andrey V. Chubukov}
\affiliation{School of Physics and Astronomy and William I. Fine Theoretical Physics Institute, University of Minnesota, Minneapolis, MN 55455, USA}

\begin{abstract}
We analyze potential non-analytic terms in the Landau diamagnetic susceptibility, $\chi_{dia}$, at a
finite temperature $T$ and/or
in-plane magnetic field $H$
in a two-dimensional (2D) Fermi liquid. To do this, we express the diamagnetic susceptibility as
$\chi_{dia}  = (e/c)^2 \lim_{Q\rightarrow0} \Pi^{JJ}_\perp (Q)/Q^2$, where
$\Pi^{JJ}_\perp$ is the transverse component of the static current-current correlator, and  evaluate
$\Pi^{JJ}_\perp (Q)$ for a system of fermions with Hubbard interaction to second order in Hubbard $U$ by combining
self energy, Maki-Thompson, and Aslamazov-Larkin diagrams.
We find that at $T=H=0$, the expansion of $\Pi^{JJ}_\perp (Q)/Q^2$ in $U$ is regular, but at a finite $T$ and/or $H$, it contains  $U^2 T$ and/or $U^2 |H|$ terms.  Similar terms have been previously found for the paramagnetic Pauli susceptibility.   We obtain the full expression for the non-analytic $\delta \chi_{dia} (H,T)$ when both $T$ and $H$ are finite, and show that the $H/T$ dependence is similar to that for the Pauli susceptibility.
\end{abstract}
\maketitle
\section{Introduction}
This communication is about the Landau diamagnetic susceptibility, $\chi_{dia}$,
of interacting electrons in a 2D Fermi liquid.
Landau diamagnetism comes from the orbital motion of electrons in the presence of a transverse magnetic
field ~\cite{landau1930diamagnetism, landau2013statistical}.
For a 2D material, the Landau diamagnetic susceptibility $\chi_{dia}$ measures the response to an infinitesimally small out-of-plane magnetic field.
For non-interacting fermions, $\chi_{dia}$  is one third in magnitude and opposite in sign to the paramagnetic Pauli susceptibility $\chi_{para}$, associated  with the alignment of the electron spin in an applied magnetic field.

The behavior of paramagnetic susceptibility is well understood.  For interacting electrons, $\chi_{para}$ at zero temperature and in the limit of zero magnetic field differs from  free-fermion expression $\chi_{para}^0 = 2 \mu_B^2 N_F$ by a factor~ \cite{lifshitz2013statistical,pinesnozieres}:
\begin{align}
\label{zero} \chi_{para} = \chi_{para}^0 \frac{m^*/m}{1+F^s_0},
\end{align}
where $m$ is the electron mass, $m^*$ is the effective mass, dressed by the interaction,  and $F^s_0$ is the Landau coefficient in the spin channel with angular momentum component $l=0$.
 Both $m^*/m$ and $F^s_0$ can be obtained perturbatively, in the expansion either in dimensionless  $r_s$ for the Coulomb interaction, or in the Hubbard $U$ for short-range interaction (the dimensionless expansion parameter is $N_F U$, where $N_F = m/(2\pi)$ is the density of states on the Fermi surface). In the Galilean-invariant case, the expansion in $N_F U$ in 2D yields, to order $(N_F U)^2$ \cite{chubukov2018fermiliquid}
 \bea
\frac{m^*}{m} &= 1 + \frac{1}{2} (N_FU)^2 \nonumber \\
 F^s_0 &= - N_F U + (N_F U)^2 \log{2} \nonumber \\
\chi_{para} &= \chi_{para}^0 \left(1 + N_F U  + (N_F U)^2 \left(\frac{3}{2} - \log{2}\right) \right).
\label{ex_1}
\eea

At a finite temperature $T$ and/or a finite magnetic field $H$, $\chi_{para}$ has been obtained
by analyzing corrections to Landau Fermi liquid theory both in 3D and in 2D~\cite{doniach1966lowtemperature,belitz1997nonanalytic,millis_1,betouras2005thermodynamics, chubukov2003nonanalytic,belitz2014nonanalyticities,shekhter2006temperature,chubukov2006galillean,drukier2015,chubukov2003nonanalytic, betouras2005thermodynamics,chubukov2003nonanalytic,maslov2006nonfermiliquid}. The Pauli susceptibility of free fermions has a regular expansion in $(T/E_F)^2$ and $(\mu_B H/E_F)^2$, where $E_F$ is the Fermi energy.
In the presence of interactions, the functional form changes:
$\chi_{para} (T, H)$ in 2D has a linear in $T$ dependence at small $T$ and a linear in $H$ dependence at small $H$. These dependencies, along with a $|Q|$ dependence of $\chi_{para} (Q)$ at $T=H=0$, come exclusively from backscattering and reflect a special  role of the subset of 1D scattering processes in a multi-dimensional system (a 2D system in our case).  More specifically, 1D scattering accounts for the $\Omega/q$  form of the Landau damping in the limit when $\Omega \ll v_F q$.  Because of the $1/q$ dependence,  the effective interaction dressed by Landau damping is long-ranged. A finite $T$ and/or a finite $H$ acts as a mass term that converts long-range interaction into a short-range one. This makes the derivatives $d \chi_{para}(T)/d T^2$ and $d \chi_{para} (H)/d H^2$ singular, if one attempts a typical power series expansion of $\chi_{para}$ in powers of  $T^2$ or $H^2$.
For this reason, we call these terms non-analytic even though $d \chi_{para}(T)/d T$ and $d \chi_{para} (H)/d H$ are finite.

For the 2D Hubbard model, the paramagnetic spin susceptibility to order $U^2$ at a finite $T$ and $H=0$  and at a finite $H$ and $T=0$ are \cite{chubukov2003nonanalytic,maslov2006nonfermiliquid}
 \bea
  \delta \chi_{para} (T) &=  \chi_{para} (T) - \chi_{para} (0) =  \chi_{para}^0 \frac{N_F^2 U^2}{2} \frac{T}{E_F}, \nonumber \\
\delta \chi_{para} (H)  &= \chi_{para}^0 N_F^2 U^2 \frac{\mu_B |H|}{E_F}
 \eea
 When both $H$ and $T$ are non-zero, we have~\cite{betouras2005thermodynamics}
\begin{align}
\delta \chi_{para} (H,T) =
\nonumber \chi_{para}^{0}  \frac{N_F^2 U^2}{2} \frac{\mu_B H}{E_F} \csch^2 \\
 \times \left( \frac{\mu_B H}{T} \right) \left[ \sinh \left( 2\frac{\mu_B H}{T} \right) - \frac{\mu_B H}{T}\right].
\label{ss_10}
\end{align}

The linear in $T$ behavior of  the paramagnetic spin susceptibility in 2D has been detected in
iron pnictides~\cite{korshunov2009nonanalytic, klingeler2010local, wang2009peculiar}.   The same physics gives rise to non-analytic temperature dependence of the specific heat coefficient, $C (T)/T = a_2 + b_2T$ in 2D and $C(T)/T = a_3 + b_3 T^2 \log T$ in 3D (see e.g., Refs. \cite{belitz1997nonanalytic,millis,glazman,millis_1}).
The latter has been observed first in $\text{UAl}_2$ \cite{trainor1975specific} and later in other uranium alloys as well as $\text{TiBe}_2$ \cite{BAKER2021153282,ANTONIO2018154,stewart1982specific}.
 The linear in $T$ behavior of $C(T)/T$ has also been observed in helium films on a variety of substrates~\footnote{ \label{helium_ref} For an alternate explanation, see R.H. Tait and J.D. Reppy, Phys. Rev. B 20, 997, (1978), as well as C.J. Yeager, L.M. Steele, and D. Finotello, Phys Rev. B 51, 15274 (1995)}.

The goal of our work is to perform the same type of analysis for the diamagnetic susceptibility, $\chi_{dia}$.  It has been argued~\cite{lifshitz2013statistical,pinesnozieres,tsvelikqft,levitov} that the Landau diamagnetic susceptibility
for interacting fermions
cannot be obtained within Fermi liquid theory as some interaction-induced corrections come from fermions away from the Fermi surface.
Still, $\chi_{dia}$ can be computed directly in the expansion in either $r_s$ or $N_F U$.  We consider short-range interaction and compute  $\chi_{dia}$ to second order in $N_F U$ in 2D. We address two issues: (i) whether $\chi_{dia} (T=H=0)$ is a regular function of $N_F U$  and (ii) whether $\chi_{dia} (T,H)$ is a non-analytic function of temperature and in-plane magnetic field (we consider the case of an infinitesimally small transverse field, which causes orbital motion of 2D fermions,  and a finite Zeeman field within the plane). It is not clear a'priori whether $\chi_{dia} (T,H)$ has to be non-analytic.  On one hand, it is a component of the magnetic susceptibility, and its counter part, $\chi_{para}$, is non-analytic. On the other hand, $\chi_{dia}$ is expressed via the correlator of charge currents (Eq. (\ref{dia}) below) and one may argue that it should have the same properties as a charge susceptibility.
The latter does not have a non-analytic $T$ and $H$ dependence because it measures the response to a variation of the chemical potential $\mu$, and such a variation preserves  $\Omega/q$ form of the Landau damping~\cite{kirkpatrick}.

Regarding these issues, we first
show that $\chi_{dia} (T=H =0)$ is regular, much like $\chi_{para}$ in Eq. (\ref{ex_1}).   The only difference is that the linear in $U$ term is absent.
Because  $\chi_{dia} (T=H =0) \propto \Pi^{JJ}_{\perp} (Q, H=T=0)/Q^2$, where $\Pi^{JJ}_{\perp} (Q, H=T=0)$ is  the transverse current-current correlator (Eq. (\ref{dia} below), a regular $\chi_{dia} (T=H =0)$  implies
$\Pi^{JJ}_{\perp} (Q, H=T=0)$ scales as $Q^2$. This is expected but not a'priori guaranteed as we will see that individual diagrams for $\Pi^{JJ}_{\perp} (Q, H=T=0)$ do contain
 $|Q|$ terms.  Such terms exist for spin-spin correlator, where they combine into a non-zero total $|Q|$ term, and for charge-charge correlator, where $|Q|$ contributions from individual diagrams cancel out. We show that for $\Pi^{JJ}_{\perp} (Q, H=T=0)$ the $|Q|$ terms from individual diagrams cancel out. In this respect, the behavior of the current-current correlator at $T=H=0$ is similar to that of the charge-charge correlator. Our analysis of $\chi_{dia} (T=H=0)$ complements several earlier studies~\cite{vignale1988diamagnetic, singhpathak, tao2006analytic}, which computed
$\chi_{dia} (T=H=0)$ for a system with Coulomb interaction.

We then discuss the nonanalyticity of $\chi_{dia}$. We show that at a finite $T$ and/or finite $H$, $\Pi^{JJ}_{\perp}$ evaluated to order $Q^2$ contains $Q^2 |H|$ and $Q^2 T$  terms, i.e., $\chi_{dia} (T, H)$ is nonanalytic, much like $\chi_{para} (T, H)$.
We re-iterate, to avoid misunderstanding, that  $\chi_{dia} (T, H)$ is the response to an infinitesimal out-of-plane magnetic field. The in-plane magnetic field only serves to induce a spin dependent dispersion via Zeeman splitting. We combine $\chi_{para}$ and $\chi_{dia}$ and obtain the non-analytic term in the full magnetic susceptibility.

In order to detect the non-analytic terms in the diamagnetic susceptibility, we recommend reexamining the measurements on the magnetic susceptibility of iron pnictides, particularly Ba$_2$Fe$_{2-x}$Co$_x$As$_2$. In earlier experiments, a magnetic susceptibility was obtained as a response to an in-plane magnetic field and had only a paramagetic (spin) component.
If a transverse magnetic field is applied, then both the spin and orbital parts should contribute to the magnetic susceptibility. Subtracting the in-plane magnetic field susceptibility from the out-of-plane susceptibility, one can potentially isolate the diamagnetic susceptibility and analyze its dependence on $T$ and $H$.
\section{General Theory}
\begin{figure}[h!]
	\begin{center}
		\includegraphics[scale=0.1]{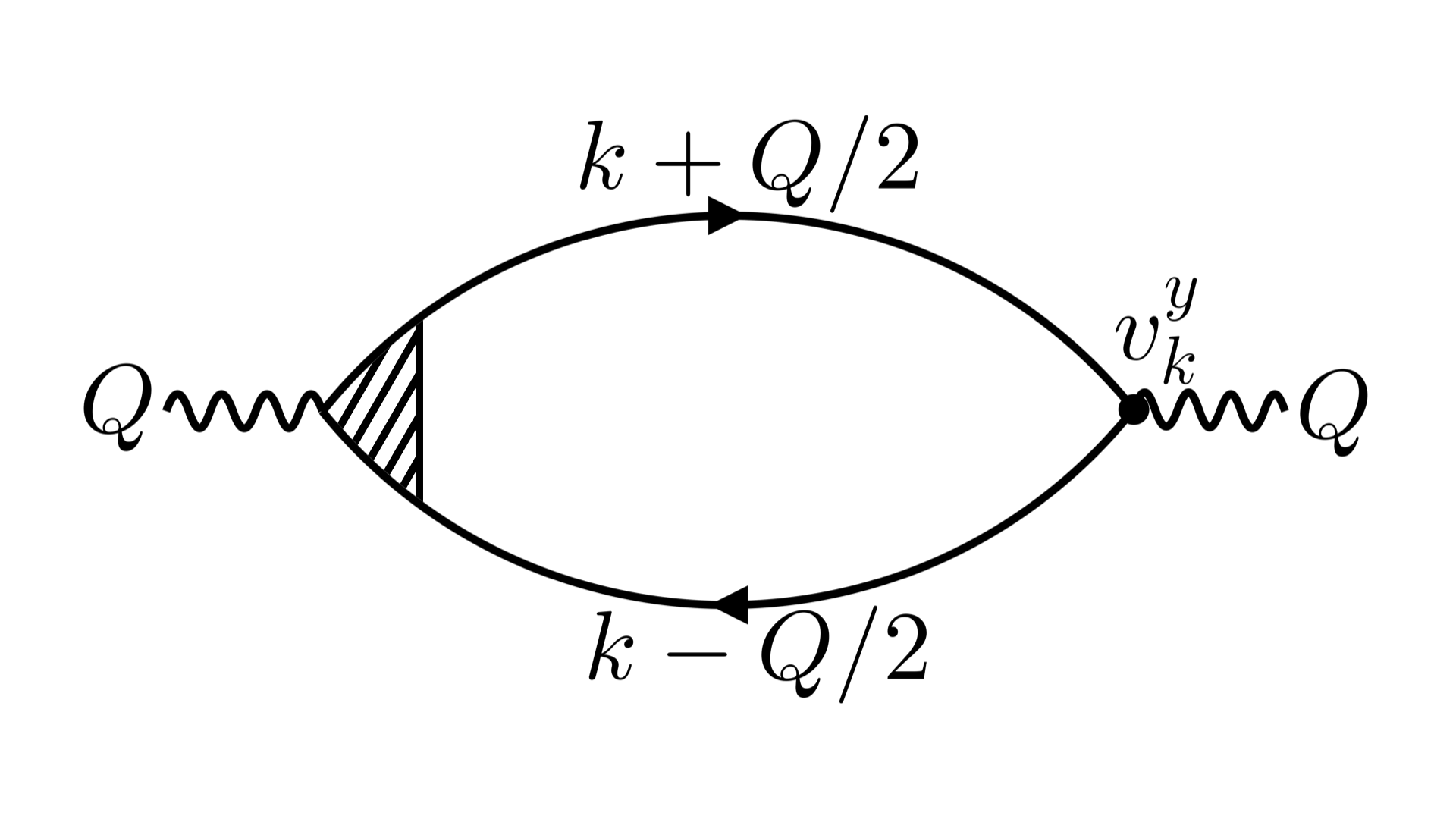}
		\caption{The fully dressed polarization bubble that represents the current-current correlation function.}
		\label{fully_dressed}
	\end{center}
\end{figure}

The static diamagnetic susceptibility is  related to the static current-current correlation function as
\begin{align}
\label{dia} \chi_{dia}  = \frac{e^2}{c^2} \lim_{Q\rightarrow0} \frac{\Pi^{JJ}_\perp (Q)}{Q^2}
 = 4 m^2 \mu^2_B \lim_{Q\rightarrow0} \frac{\Pi^{JJ}_\perp (Q)}{Q^2},
\end{align}
where $\Pi^{JJ}_\perp$ is the component of the current-current correlation perpendicular to the direction of $Q$ \cite{vignale1988diamagnetic, pinesnozieres}
 (we set $\hbar =1$).
 This $\Pi^{JJ}_\perp (Q)$ is the total current-current correlator, subject $\lim_{Q\rightarrow0}\Pi^{JJ}_\perp(Q,0) = 0$~\cite{giuliani2005quantum}. Diagrammatically, $\Pi^{JJ}_\perp (Q)$ is expressed as the fully dressed particle-hole bubble with full Green's functions and one dressed and one bare current vertex (Fig. \ref{fully_dressed}).  In analytic form,
\begin{align}
\Pi^{JJ}_{\perp}(\vb Q) = -2 T \sum_{\omega_m} \int \frac{d^2k}{(2\pi)^2} v_{\vb k}^{\perp} \Gamma_{\perp} ({\vb k}, {\vb Q})
 \left(G_Q - G_{Q \to 0}\right),
\label{ex_22}
\end{align}
where $v_{\vb k}{\perp}$ is the  component of the velocity perpendicular to the direction of $\vb Q$, $\Gamma_{\perp} ({\vb k}, {\vb Q})$ is the fully dressed transverse  current vertex, and $G_Q =  G(\vb k + \vb Q/2, \omega_m) G(\vb k - \vb Q/2, \omega_m)$.

For non-interacting fermions,
$\Gamma_{\perp} ({\vb k}) =v_{\vb k}^{\perp}$, and
\begin{align}
\nonumber \Pi^{JJ}_{\perp}(\vb Q) = -2 T \sum_{\omega_m} \int \frac{d^2k}{(2\pi)^2} \left( v_{\vb k}^{\perp} \right)^2
\big(G_0(\vb k + \vb Q/2, \omega_m) \times \\
G_0(\vb k - \vb Q/2, \omega_m) - G^2_0 (\vb k, \omega_m)\big),
\label{ex_2}
\end{align}
where  $G_0(\vb k, \omega_m) = \left( i \omega_m - \e_k\right)^{-1}$ is free-fermion Green's function.  At $T=0$,
$T \sum_{\omega_m} = \int d\omega_m/(2\pi)$.   The momentum and frequency integral is infra-red and ultra-violet convergent and can be evaluating by integrating over momentum and frequency in any order.
For a parabolic dispersion, Eq. (\ref{ex_2})
 yields, to lowest order in $Q$, $\Pi^{JJ}_{\perp}(\vb Q) = - Q^2N_F/(6 m^2)$ in both 2D and 3D, and Eq. (\ref{dia})
 reproduces the usual expression for the Landau diamagnetic susceptibility, $\chi_{dia}^0 = -\frac{1}{3} \chi_{para}^0$
\cite{lifshitz2013statistical,pinesnozieres}.
We show in Appendix \ref{landaupeierls} that for a single band with an arbitrary dispersion, expandable in even powers of momentum, Eq. (\ref{ex_2}) reproduces the Landau-Peierls expression:
\begin{align}
\chi_{dia} = \frac{2\mu_B^2 m^2}{3(2 \pi)^d } \int d^dk \, n_F'(\e_k) \left( \frac{\partial^2 \e_{\vb k }}{\partial k_x^2}  \frac{\partial^2 \e_{\vb k }}{\partial k_y^2} - \left( \frac{ \partial^2 \e_{\vb k}}{\partial k_x \partial k_y}\right)^2\right).
\end{align}

\begin{figure}[h]
	\begin{center}
		\includegraphics[scale=0.03]{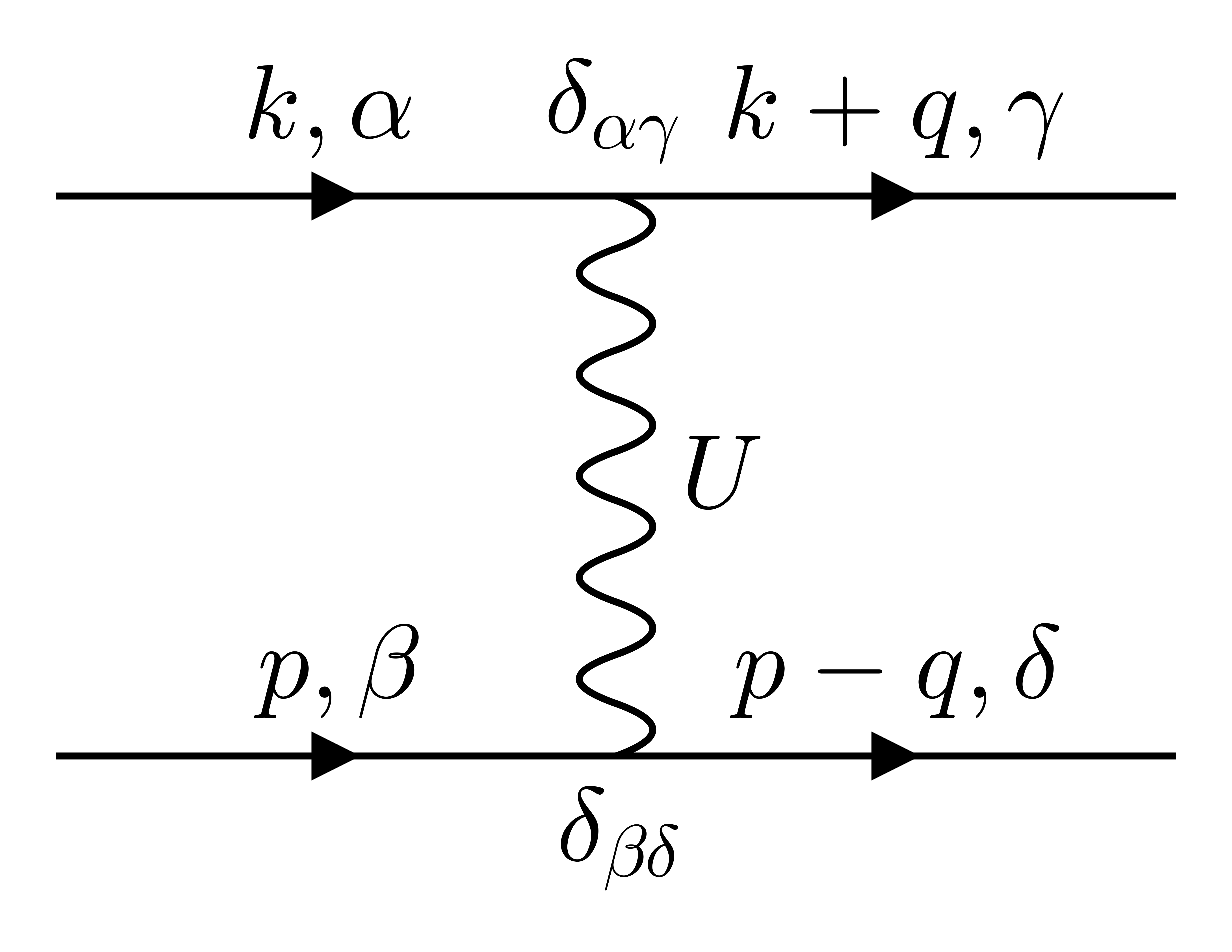}
		\caption{The vertex function in the Hubbard model to first order in $U$.}
		\label{hubbard_vertex}
	\end{center}
\end{figure}

 The key interest of our study is the diamagnetic  susceptibility for interacting fermions both at $T=H=0$ and when either temperature or an in-plane field (or both) are finite.  To make computation less involved, we consider Hubbard interaction between fermions and assume that fermion dispersion is parabolic. We present the Hubbard vertex in Fig. \ref{hubbard_vertex}.

 The full $\Pi^{JJ}_{\perp}(\vb Q)$ for an interacting system is obtained by adding vertex corrections to the free-fermion bubble and by dressing
the fermionic propagators.   Diagrams for $\Pi^{JJ}_{\perp}(\vb Q)$  to first and second order in $N_F U$ are presented in Figs \ref{first_order} and \ref{secondorder}. The wavy lines in these diagrams are the Hubbard $U$.
In Fig. \ref{first_order}, the diagram with the renormalization of the fermionic line is traditionally called the ``self-energy" or ``density of states" diagram and the one with the vertical wave line is called Maki-Thompson diagram.  In Fig. \ref{secondorder} for $\Pi^{JJ}_{\perp}(\vb Q)$ to order $U^2$, the first two diagrams renormalize $G$ into $G_0$, others renormalize  one $v_{\vb k}^{\perp}$ into $\Gamma_{\perp} ({\vb k}, {\vb Q})$. The last two  diagrams in Fig. \ref{secondorder} (diagrams f and g) are traditionally called Aslamazov-Larkin diagrams and we will use this notation
 \footnote{
 Historically,  Aslamazov-Larkin diagrams have been introduced to account for  contributions from fluctuations in the particle-particle channel (A. Larkin and A. Varlamov, \textit{Theory of Fluctuations in Superconductors}, Vol. 127 (OUP Oxford, 2005)). In the context of fluctuation corrections in the particle-hole channel, due to Coulomb interaction, these diagrams have been introduced by  Ma and Bruekner (S-K Ma and K. A.  Brueckner, Phys. Rev. 165, 18 (1968)). We thank D.L. Maslov for clarifying this issue.}.

We will see below that the full $\Pi^{JJ}_{\perp}(\vb Q)$ at order $U^2$ and at $T=0$, $H=0$ can be expressed in terms of the three diagrams (a), (c), and (g) in Fig. \ref{secondorder} (other four diagrams in Fig. \ref{secondorder} are expressed in terms on these three).  To make our notation more concise, we will designate the diagram (a) as the second-order self-energy diagram, diagram (b) as the second order Maki-Thompson diagram, and diagram (g) as the Aslamazov-Larkin diagram.

\section{Nonanalyticities of the Polarization Bubble}

Diagrams (a), (c) and (f)
in Fig. \ref{secondorder} all contain
a polarization bubble of free fermions $\Pi_{ph} (q, \Omega_m)$.
Before we proceed with the calculation of these diagrams first at $T=H=0$ and then at finite $T$ and $H$, it is instructive to list the expressions for
$\Pi_{ph} (q, \Omega_m)$ at small frequency $\Omega_m$ and momenta $q$ near either $0$ or $2k_F$, as these expressions will determine the non-analyticities of $\Pi^{JJ}_{\perp}(\vb Q) (T, H)$~ \cite{carneiro1977susceptibility, belitz1997nonanalytic, chubukov2003nonanalytic, betouras2005thermodynamics,chitov2001leading,betouras2005thermodynamics}.

At $T=H=0$, the particle-hole polarization bubble of free fermions in 2D is given by
\begin{align}
\Pi_{ph} (q, \Omega_m) = \int \frac{d^2 k}{(2\pi)^2} \frac{d \omega_n}{2\pi} G(\vb k,\omega_n) G(\vb k+ \vb q,\omega_n+\Omega_m),
\end{align}
At small $q$ and $\Omega_m$,
\begin{align}
\Pi^{q\rightarrow0}_{ph}(q,\Omega_m) = \frac{m}{2\pi} \left( -1 +  \frac{|\Omega_m|}{\sqrt{\left(v_F q \right)^2 + \Omega^2_m}}\right).
\label{ex_5}
\end{align}
At $v_F q \gg |\Omega_m|$, $\Pi^{q\rightarrow0}_{ph}(q,\Omega_m)$ contains a non-analytic $|\Omega_m|/q$  term.

Near $q= 2k_F$
\begin{align}
\nonumber &\Pi^{q \rightarrow 2 k_F}_{ph} (q,\Omega_m) \\
&=\frac{m}{2\pi} \left( -1 + \frac{1}{2} \left( \sqrt{ \frac{ \tilde q}{k_F} - \frac{i \Omega_m}{v_F k_F}} + \sqrt{ \frac{ \tilde q}{k_F} + \frac{i \Omega_m}{v_F k_F}} \right) \right),
\end{align}
where $\tilde q = q - 2k_F$.  For ${\tilde q} <0$ and $v_F |\tilde q| > |\Omega_m|$, $\Pi^{q \rightarrow 2 k_F}_{ph} (q,\Omega_m)$ again contains a non-analytic $|\Omega_m|/{\tilde q}$  term, as one can readily verify by expanding in small $\Omega_m/{\tilde q}$ around the branch cuts in the square roots.

These forms of the polarization bubbles  give rise to the appearance of  non-analytic $|Q|$ terms in the individual diagrams for the current-current correlation function already at $T=H=0$.  We show later that these nonanalyticities cancel once all diagrams are added together.

For the analysis of non-analyticities in $\chi_{dia}$, we will need the expressions of the polarization bubble at a finite $T$ and/or in-plane $H$.
When $T=0$ and $H$ is finite, the polarization is  spin dependent:
\begin{align}
\Pi^{\alpha \beta}_{ph}(q,\Omega_m) = \int \frac{d^d k}{(2\pi)^d} \frac{d \omega}{2\pi} G^{\alpha}(\vb k,\omega) G^{\beta}(\vb k+ \vb q,\omega+\Omega_m)
\end{align}
Then one has to distinguish between $\Pi^{\uparrow \uparrow }_{ph} (q, \Omega_m)$ and $\Pi^{\uparrow \downarrow}_{ph}(q, \Omega_m)$. At small $q$ and $\Omega_m =0$,
\begin{align}
\Pi^{\uparrow \uparrow }_{ph} (q, \Omega_m) &= \frac{m}{2\pi}\frac{|\Omega_m|}{v_F q}
+ \cdots \\
\Pi^{\uparrow \downarrow}_{ph} (q, \Omega_m) &= \frac{m}{2\pi}\frac{|\Omega_m|}{\sqrt{\left(v_F q \right)^2 - (2 \mu_B H)^2}}
+ \cdots
\end{align}
 where dots stand for regular terms.
 We see that a finite $H$ is crucial for $\Pi^{\uparrow \downarrow}_{ph} (q, \Omega_m)$, where it cuts a long-range interaction and causes singularity in the derivative with respect to $H$, but not essential for $\Pi^{\uparrow \uparrow}_{ph} (q, \Omega_m)$.
 Near $q=2k_F$, the situation is opposite:
\begin{align}
\nonumber \Pi^{\uparrow \uparrow}_{ph}(q,\Omega_m) = &\frac{m}{4\pi} \Bigg( \sqrt{ \frac{ \tilde q}{k_F} - \frac{i \Omega_m}{v_F k_F}-\frac{2\mu_B H}{v_F k_F}} \\ 
&+ \sqrt{ \frac{ \tilde q}{k_F} + \frac{i \Omega_m}{v_F k_F} - \frac{2 \mu_B H}{v_F k_F}} \Bigg) + \cdots\\
\Pi^{\uparrow \downarrow}_{ph}(q,\Omega_m) = & \frac{m}{4\pi}  \left( \sqrt{ \frac{ \tilde q}{k_F} - \frac{i \Omega_m}{v_F k_F}} + \sqrt{ \frac{ \tilde q}{k_F} + \frac{i \Omega_m}{v_F k_F}} \right) + \cdots,
\end{align}
where the ellipses again stand for analytic terms. We see that a finite $H$ affects  the term where both spin indices are the same and does not affect the term with opposite spin indices.
Below we combine fermions into particle-hole pairs in such a way that we only get terms  $\Pi^{\uparrow \downarrow}_{ph}(q,\Omega_m)$. With this  we ensure that all nonanalytic contributions come from only internal $q\approx 0$.

At a finite $T$ and $H=0$, the particle-hole polarization bubble near $q=0$  has the same form  as at $T=0$, Eq. (\ref{ex_5}), but now Matsubara frequencies are discrete, $\Omega_m = 2\pi mT$. The dynamical piece is present at $m \neq 0$, when $\Omega_m$ are finite. The same finite $\Omega_m$ then appears in the denominator and cuts  long-range interaction at $v_F q < 2 \pi T$, i.e., at distances $r > v_F/(2\pi T)$.  This in turn causes singularity in the temperature derivative of $\Pi_{ph} (q, \Omega_m)$.

We are not aware of a closed form of the polarization bubble near $q=2k_F$ at finite temperature.
 In Appendix \ref{appT2kf} we compute the linear in $T$ contribution to $\chi_{dia} (T)$ from momenta near $2k_F$ directly, without expressing it via the $2k_F$ polarization bubble, and for completeness also do the same for the contribution from $q$ near zero.

\section{Zero Temperature and Zero Magnetic Field}
\subsection{First Order in $U$}
\begin{figure}[h]
	\begin{center}
		\includegraphics[scale=.14]{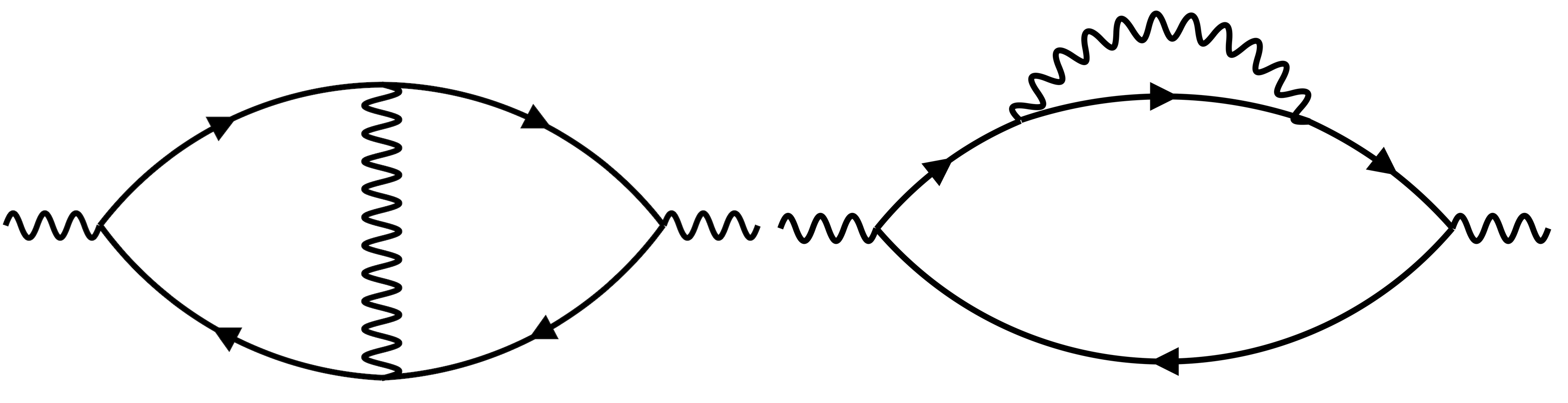}
		\caption{The two distinct diagrams which appear at first order in $U$. The first diagram is
 often called the Maki-Thompson diagram and the second is often called
 the self energy diagram.}
		\label{first_order}
	\end{center}
\end{figure}

We first consider corrections to the diamagnetic susceptibility in the Hubbard model at both $T=0$ and $H=0$.
The diagrams for $\Pi^{JJ}_{\perp} (Q,0)$ are shown in Fig. \ref{first_order}. These diagrams have already been evaluated for the diamagnetic susceptibility in the case of a dynamically screened Coulomb interaction in RPA \cite{vignale1988diamagnetic}. We show that in the Hubbard model, each of these diagrams evaluate to 0.
\\
We can write the contribution of the Maki-Thompson diagram as
\begin{align}
\Pi^{JJ,MT}_{\perp}(Q,0) = 2 U \left( \int \frac{d^2k}{\left(2\pi\right)^2} \frac{d\omega}{2\pi} v_k^y G_{k-Q/2} G_{k+Q/2} \right)^2,
\end{align}
where $G_{k \pm Q/2} = G(\vb k \pm \vb{Q} /2, \omega)$. Taking $\vb k \rightarrow - \vb k$, and noting $\e_{-\vb k} = \e_{\vb k}$, we find
\begin{align}
\nonumber \int &\frac{d^2k}{\left(2\pi\right)^2} \frac{d\omega}{2\pi} v_k^y G_{k-Q/2} G_{k+Q/2}  \\
&= - \int \frac{d^2k}{\left(2\pi\right)^2} \frac{d\omega}{2\pi} v_k^y G_{k-Q/2} G_{k+Q/2}  = 0.
\end{align}
For the self energy diagram, we first note that there is a combinatorial factor of two in addition to the factor of two due to spin summation. The resulting susceptibility is then
\begin{align}
\nonumber \Pi_{\perp}^{JJ,SE}(Q,0) = 4 U \left(\int \frac{d^2k'}{(2\pi)^2} \frac{d \omega'}{2\pi} G_{k'} \right)\\
\times \left( \int \frac{d^2 k}{(2\pi)^2} \frac{d \omega}{2\pi} \left( v_k^y \right)^2 G_{k+Q/2}^2G_{k-Q/2} \right)
\end{align}
 Changing $\vb k \rightarrow - \vb k$ in the second term, we find
 \begin{align}
\nonumber &\int d^2k d \omega (v_k^y)^2 G_{k+Q/2}^2 G_{k-Q/2} \\
= &\int d^2k d \omega (v_k^y)^2 G_{k+Q/2} G_{k-Q/2}^2.
\end{align}
On the other hand, doing frequency integration first, we find
\begin{align}
\nonumber \int d \omega \, G_{k+Q/2}^2G_{k-Q/2} &= i \int d \omega \frac{ \partial}{ \partial \omega } \left(G_{k+Q/2} \right) G_{k-Q/2}\\
\nonumber &= - i \int d \omega  G_{k+Q/2} \frac{ \partial}{ \partial \omega } \left(G_{k-Q/2} \right) \\
&= - \int d \omega \, G_{k+Q/2} G_{k-Q/2}^2
\end{align}
Comparing the two expressions, we see that $\Pi_{\perp}^{JJ,SE}(Q,0) =0$. We then must move to second order in $U$ to detect the effects of interaction.
\begin{figure}[h]
	\begin{center}
		\includegraphics[scale=.17]{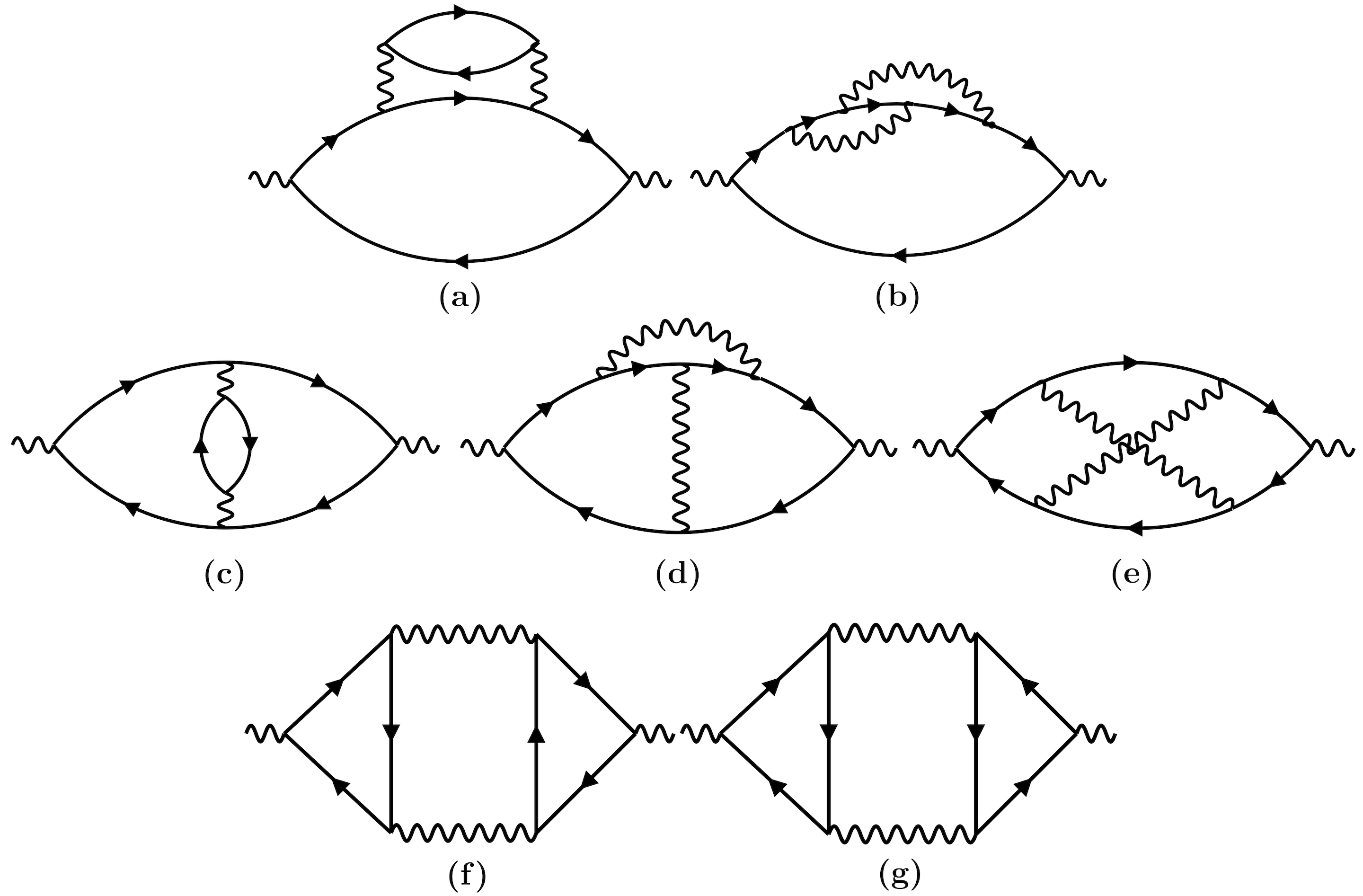}
		\caption{The seven diagrams which contribute to the current-current correlator at order $U^2$. We call diagram (a) the second order self energy correction, diagram (c) the second order Maki-Thompson correction, and (g) the Aslamazov-Larkin diagram.}
		\label{secondorder}
	\end{center}
\end{figure}
\subsection{Second Order in $U$}
To second order in $U$, there are a total of seven nontrivial diagrams that contribute to the current-current correlator, as shown in Fig. \ref{secondorder}.
We call the corresponding contribution $\Pi_i$ ($i=a$ to $g$).
We incorporate factors of 2 from combinatorics and from spin summation into $\Pi_i$.

The calculation of the diagrams is tedious but straightforward. We present some details in Appendices \ref{nonanalyticity} and \ref{ap1} and here cite the results.  First, we verified that there are particular relations between different  $\Pi_i$, namely $\Pi_{a} = - 2 \Pi_{b}$, $\Pi_{c} = \Pi_{f} = - \Pi_{d}$, and $\Pi_{e} = -\frac{1}{2} \Pi_{g}$. The total contribution will then be
\begin{align}
\nonumber \Pi^{JJ}_{\perp}(Q,0) &=\frac{1}{2} \Pi_{a}  + \Pi_{c} + \frac{1}{2} \Pi_{g}\\
&= \frac{1}{2} \left(\Pi_{a}  + \Pi_{c}\right) + \frac{1}{2} \left(\Pi_f + \Pi_g\right)
\end{align}
Next, we find that $O(Q^2)$ contributions from diagrams (f) and (g) cancel (see Appendix \ref{ap1}).
Then we can write the current-current correlator as
\begin{align}
\nonumber \Pi^{JJ}_{\perp}(Q,0) &= \frac{1}{2} \left(\Pi_{a}  + \Pi_{c}\right) = 2 U^2 \int_{k,q} \Pi(q,\Omega_m)\\ 
\nonumber &\times \Big(2 \left(v_k^y\right)^2 G_{k+Q/2}^2 G_{k-Q/2} G_{k+q+Q/2}  \\
 \label{u_2}  &+ v_k^y v_{k+q}^y G_{k+Q/2} G_{k+q+Q/2} G_{k-Q/2} G_{k+q-Q/2} \Big),
\end{align}
where $\Pi(q,\Omega_m) = \int \frac{d^2p}{(2\pi)^2} \frac{\omega_p}{2\pi} G_p G_{p+q}$ and we have used the abbreviation $\int_{k} = \int d^2k d\omega_n/(2\pi)^3$.
Finally, we verified that while  both $\Pi_a$ and $\Pi_c$ contain non-analytic $|Q|$ terms, the sum of the two has no net nonanalyticity, i.e., the expansion in $Q$ starts with $Q^2$, the details of which are in Appendix \ref{nonanalyticity}

In explicit calculation of $\Pi^{JJ}_{\perp}(Q,0)$ from (\ref{u_2}), we evaluate the integral over $\omega_n$ first, then expand out to order $Q^2$. This procedure ensures that relevant contributions from $\omega_m \sim v_F Q$ are all included.
The result is
\begin{widetext}
\begin{align}
\label{numericintegral} \Pi^{JJ}_{\perp} (Q,0) = -\frac{m Q^2 U^2 }{(2\pi)^4} \int \limits_{0}^\infty d\Omega_m \int \limits_0^\infty \frac{dq}{6 q^3} \left( \sqrt{\alpha^2 -1} + \sqrt{ (\alpha^*)^2-1} \right)  \left(\frac{\alpha ^2 q^2+2 q^2+6 \alpha q+3}{ \left(\alpha ^2-1\right)^{5/2}}+\frac{\left(\alpha ^*\right)^2 q^2+2 q^2+6 \alpha ^* q+3}{\left(\left(\alpha ^*\right)^2-1\right)^{5/2}}\right),
\end{align}
\end{widetext}
where $\alpha = \frac{i \Omega_m}{q} - \frac{q}{2}$.
Numerical evaluation of the integral gives
\begin{align}
\delta \Pi^{JJ}_{\perp} (Q,0) = - 0.2618 \frac{m Q^2 U^2 }{(2\pi)^4}
\end{align}
For the correction to the diamagnetic susceptibility, we then have
\begin{align}
 \delta \chi_{dia} &= \frac{e^2}{c^2} \lim_{Q\rightarrow 0} \frac{\delta \Pi^{JJ}_{\perp}(Q,0)}{Q^2} =
 A \chi_{dia}^0 N_F^2 U^2
\end{align}
 where
 $A = 0.7854/\pi$.
To five significant digits, $A =1/4$, and we believe that this is the exact value of $A$.

\onecolumngrid
\begin{figure*}[t]
	\begin{center}
		\includegraphics[scale=.30]{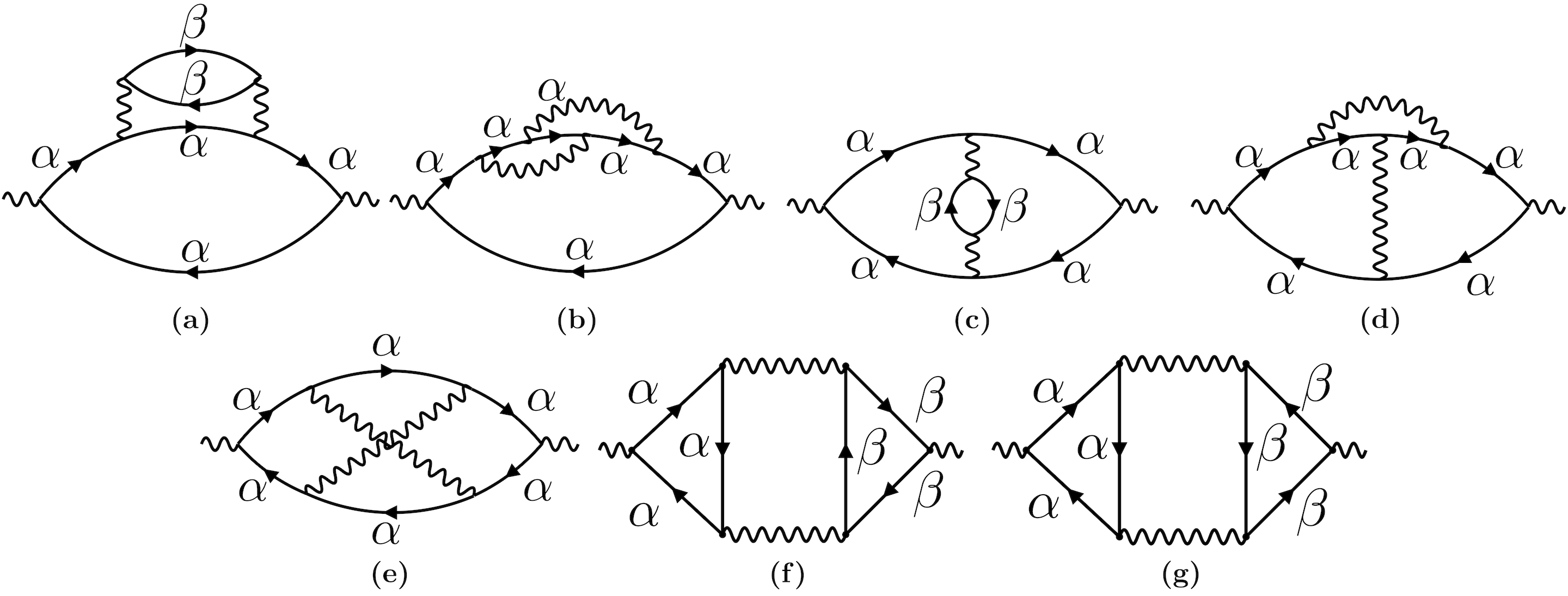}
		\caption{The relevant diagrams for the current-current correlation in the presence of a finite magnetic field. $\alpha$ and $\beta$ label spin-up and spin-down states, respectively.}
		\label{magnetic_diagrams}
	\end{center}
\end{figure*}
\twocolumngrid

 We see this correction enhances the diamagnetic susceptibility compared to that for free fermions.
We note that the sign of this correction is opposite the sign found in previous work in the case of the dynamically screened Coloumb interaction~\cite{vignale1988diamagnetic,tao2006analytic}. In that case, interactions have been found to decrease the magnitude of the diamagnetic susceptibility. However, Refs. \cite{vignale1988diamagnetic,tao2006analytic} only considered diagrams to first order in the interaction. In our case, a non-zero result for $\delta \chi_{dia}$ appears at second order in $U$.
By magnitude, $\delta \chi_{dia}/\chi^{0}_{dia}$ is about a third of $\delta \chi_{para}/\chi^{0}_{para}$ in Eq. (\ref{ex_1}). We see that even though diamagnetism is enhanced by the Hubbard interaction, the enhancement is smaller than the increase in the paramagnetic susceptibility.

 We also note that the convergence of the double integral in Eq. (\ref{numericintegral}) at large $\Omega_m$ and $q$
 implies that to second order in $U$ the diamagnetic susceptibility still comes exclusively from fermions near the Fermi surface, where one can linearize the fermionic dispersion in $k-k_F$.  At higher orders in $U$, we expect that there will be some dependence of the upper cutoff of the low-energy theory.

\section{Nonanalytic Contributions to the Current-Current Correlator}
\label{nonanalytic_h_t}

We now consider the effects of finite temperature and finite in-plane magnetic field. To do so, we consider precisely the same $U^2$ terms as before, but with either
a finite in-plane field $H$ or finite temperature $T$.
 At a non-zero $H$,  fermionic dispersion becomes spin dependent $\e_k \rightarrow \e^{\uparrow (\downarrow)}_{k} = \e_k \pm \mu_B H$, while for finite temperature the integrals over frequency are replaced with sums $\Omega_m \rightarrow 2 \pi m T$ and $\omega_n \rightarrow (2n +1) \pi T$. We recall that we have chosen an in-plane field to make a direct comparison to the case of spin susceptibility, for which a Zeeman field gives rise to non-analyticity.
We emphasize again that the Landau diamagnetic susceptibility we calculate is not the response to the Zeeman field. Rather, we are calculating how the response to an infinitesimal out-of-plane magnetic field changes when there is a finite in-plane field present.

We calculate both $\delta \chi_{dia} (H,0)$ and
$\delta \chi_{dia} (0,T)$ analytically by restricting to contributions from small  $q$ in the polarization bubble $\Pi_{ph} (q, \Omega_m)$. For a finite $H$, we argue that this is the full non-analytic contribution.  For a non-zero $T$ and $H=0$, we show that there is also the contribution from
$q \approx 2k_F$.

As a side remark, we  note that, though we only consider constant $U$ here, the calculations can be straightforwardly extended to a momentum dependent interaction $U(q)$ using the same strategy as in Ref. \cite{chubukov2003nonanalytic} for the spin susceptibility. We expect that, like in that case, the prefactors for the non-analytic terms are expressed
in terms of $U(0)^2$, $U(2k_F)^2$, or $U(0)U(2k_F)$.

\subsection{Magnetic Field}
\begin{figure}[h!]
	\begin{center}
		\includegraphics[scale=.6]{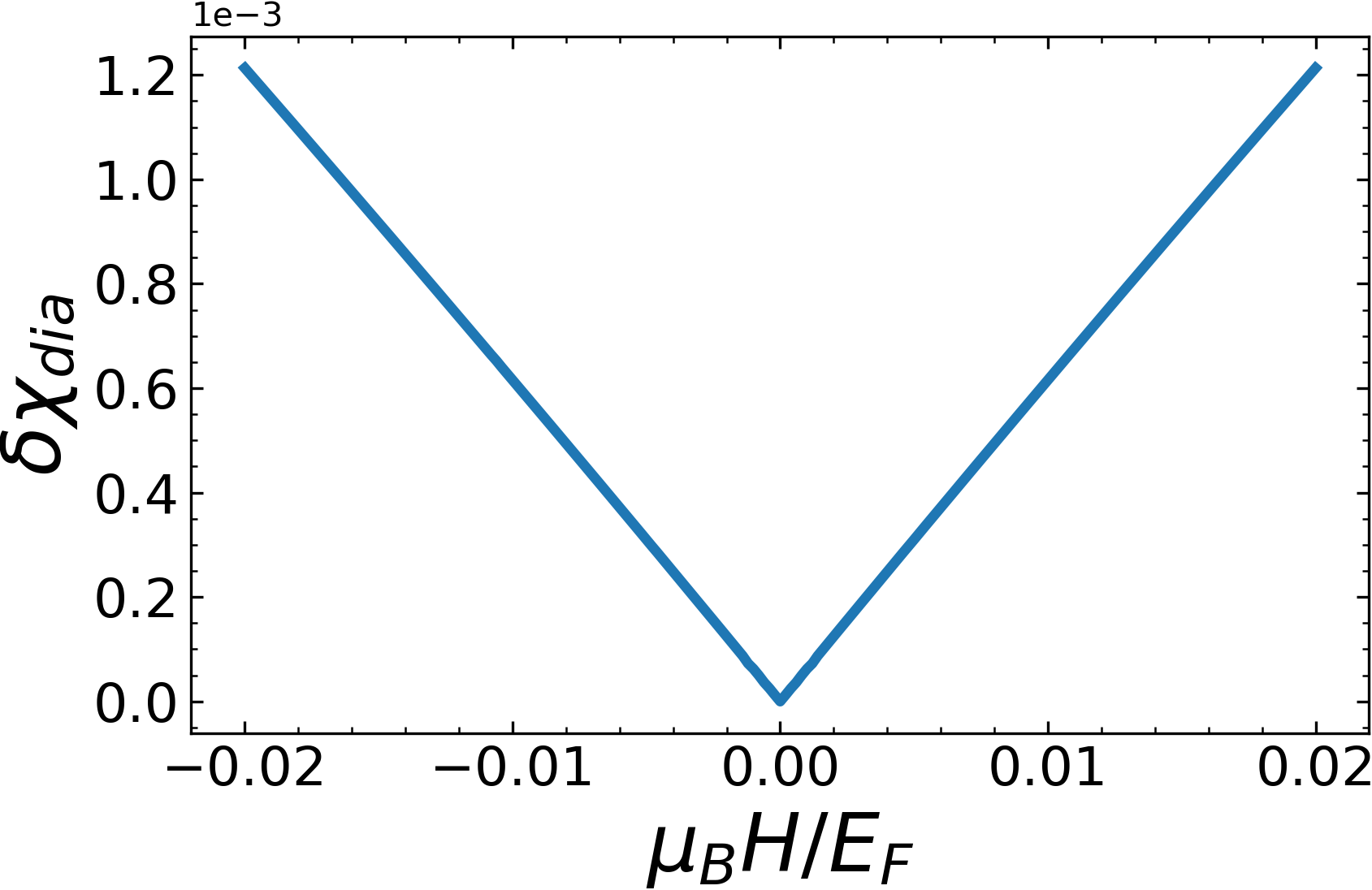}
		\caption{The numerical evaluation of diagrams (a) and (f). Here we have not restricted the magnitude of $q$ to small values. The result for $\delta \chi_{dia} (H,0) \propto |H|$ agrees with the analytical calculation done by restricting to small $q$.}
		\label{finite_H}
	\end{center}
\end{figure}

As we said, in a finite in-plane field, fermionic Green's functions become spin-dependent, $G_{k,\alpha} = (i \omega_n - \e_k^{\alpha})^{-1}$ and $\e_k^{\uparrow (\downarrow)} = \e_k \pm \mu_B H$. We first note that upon adding all diagrams, the terms that exclusively contain $G_\uparrow$ or $G_\downarrow$ will cancel. We can see this by explicitly adding up diagrams with the same momentum labeling, then noting the difference in the spin indices for each of the Green's functions.  As an example, consider diagrams (a) and (b). Writing them together, we have
\begin{align}
\nonumber&\int_{k,q} \left(v_k^y\right)^2\Bigg( \sum_{\alpha,\beta} \Pi^{\alpha \beta} (q, \Omega)  \left(G_{k+Q/2,\alpha} \right)^2 G_{k-Q/2,\alpha} G_{k+q+Q/2,\beta} \\
\nonumber &- \sum_{\alpha } \Pi^{\alpha \alpha} (q, \Omega)\left( G_{k+Q/2,\alpha} \right)^2 G_{k-Q/2,\alpha} G_{k+q+Q/2,\alpha} \Bigg)\\
=&\int_{k,q} \left(v_k^y\right)^2 \sum_{\alpha \neq \beta} \Pi^{\alpha \beta}  \left(G_{k+Q/2,\alpha} \right)^2 G_{k-Q/2,\alpha} G_{k+q+Q/2,\beta}.
\end{align}
 This immediately implies that out of seven diagrams in Fig. \ref{magnetic_diagrams}, only diagrams (a), (c), (f) and (g) contribute, with spin index $\beta \neq \alpha$. Next, we explicitly verify (see Appendix \ref{ap1}) that at order $Q^2$, diagrams (c) and (g) cancel each other, i.e., $\delta \chi_{dia} (H,0)$ is the sum of diagrams (a) and (f).  Finally, we use the fact that non-analyticity in the polarization bubble made of fermions with opposite spin projections comes from momenta $q \approx 0$ and construct the polarization bubble in the diagram (f) out of fermionic propagators shown by vertical lines (they have opposite spins $\alpha$ and $\beta$), and construct the
  polarization bubble in the diagram (a) using one of the two $\beta$ fermions and the $\alpha$ fermion ``located"
  immediately below $\beta$ fermions in Fig. \ref{magnetic_diagrams}. The sum of diagrams (a) and (f) is then expressed as
\begin{align}
\nonumber \delta \Pi^{JJ}_\perp(H)=&U^2\sum_{\alpha \neq \beta} \int_{ k,  q}  \Big[\Pi^{\alpha \beta} (q,\Omega_m) \Big( 2 (v_k^y)^2 \left(G^{\alpha}_{k+Q/2}\right)^2\\
& \times G^{\alpha}_{k-Q/2} G^{\beta}_{k+q+Q/2} + v_k^y v_{k+q}^y G^{\alpha}_{k+Q/2} \nonumber \\
 &\times G^{\alpha}_{k-Q/2} G^{\beta}_{k+q+Q/2}G^{\beta}_{k+q-Q/2}\Big)\Big]
\label{finiteT1}
\end{align}
The evaluation of this expression is again tedious but straightforward. We present the details in Appendix \ref{magfield}.
The result is
\begin{align}
\nonumber \delta \Pi^{JJ}_\perp (H) = &-\frac{U^2 Q^2 k_F^2 }{m^2 (2\pi)^4} \int_{-\infty}^{\infty}d \Omega_m \\
\label{magfieldresult} &\times \int_{0}^{\infty} dq  \frac{q^3 \Omega_m ^2 \left(4 \left(i \Omega_m + 2 \mu_B H \right)^2+v_F^2 q^2\right)}{4 \left(\left(i \Omega_m+ 2 \mu_B H \right)^2-v_F^2q^2\right){}^4}.
\end{align}
Integrating over $q$, subtracting off the $H=0$ case, and then integrating over $\Omega_m$, we find
\begin{align}
\nonumber \delta \Pi^{JJ}_{\perp}(H) &=  -\frac{U^2 Q^2 m}{24(2 \pi)^3} \frac{\mu_B |H|}{E_F}
\\ \implies \delta \chi_{dia}(H) &= \frac{1}{4}\chi^0_{dia} U^2 N_F^2 \frac{\mu_B |H|}{E_F}.
\end{align}
To verify this result, we computed the sum of diagrams (a) and (f) numerically, not restricting to small $q$. We plot the results in Fig \ref{finite_H}.  We see that there is a fairly good agreement with the analytical analysis, in which we restricted to only small $q$. The agreement confirms that non-analytic $\delta \chi_{dia} (H,0)$  comes from only $q \ll k_F$.
\subsection{Finite Temperature}
We now perform the same analysis as above in the case of $H=0$ but $T\neq 0$. We first consider analytically the contribution from small $q$, i.e. from $v_F q \sim \Omega_n \sim T$.
The calculation is very similar to the one in the previous section and the result is Eq. (\ref{finiteT1}) with $H=0$, and $\int d\Omega_m/2\pi \rightarrow T \sum_{\Omega_m}$.
Using the expression for $\Pi(q, \Omega_m)$ in Eq. (\ref{ex_5}), this equation can be re-expressed as
\begin{align}
\label{smallqT} \delta \Pi^{JJ}_{\perp,q=0} (T) = -\frac{U^2 Q^2k_F^2}{m^2(2\pi)^3} T \int dq \sum_m \frac{q^3 \Omega_m^2 \left(v_F^2 q^2-4\Omega_m^2\right)}{4 \left(\Omega_m^2+v_F^2 q^2\right){}^4},
\end{align}
We now sum over $\Omega_m$, subtract off the $T=0$ contribution, and integrate over $q$. Doing so, we find
\begin{align}
\delta \Pi^{JJ}_{\perp,q=0} (T) &= - \frac{U^2 Q^2 m}{48(2\pi)^3} \frac{T}{E_F}
\\ \implies \delta \chi_{dia}^{q=0}(T) &= \frac{1}{8}\chi^0_{dia} U^2 N_F^2 \frac{T}{E_F},
\label{nnn_1}
\end{align}
where, we recall, $\chi^0_{dia} = -\frac{2}{3} \mu_B^2 N_F$ is the bare diamagnetic susceptibility. We see that $\delta \Pi^{JJ}_{\perp,q=0} (T)$, and hence $\delta \chi_{dia}^{q=0} (0, T)$, scales linearly with $T$. For completeness, in Appendix \ref{numericT} we calculate this term by summing over the two fermionic Matsubara frequencies first, expanding to order $Q^2$, and evaluating the resulting term. The result gives precisely the same expression for this linear in T term as above.

We also analyze the linear in $T$ contribution to $\delta \Pi^{JJ}_{\perp} (T)$ from $q \sim 2k_F$.
This analysis requires more efforts, and we present it in
Appendix \ref{appT2kf}. The result is that there is a linear in $T$ contribution to $\chi_{dia}$  from $q=2k_F$,
 which is equal to the contribution at $q=0$, i.e.
\begin{align}
\delta \chi_{dia}^{q=2k_F}(T) = \frac{1}{8}\chi^0_{dia} U^2 N_F^2 \frac{T}{E_F}
\end{align}
Adding the two terms together, we find for the total nonanalytic contribution at finite temperature
\begin{align}
\delta \chi_{dia}(T) = \frac{1}{4}\chi^0_{dia} U^2 N_F^2 \frac{T}{E_F}
\end{align}

\subsection{Finite Magnetic Field and Temperature}
We note that, when internal $q\sim 0$, we do not need to set either $H=0$ or $T=0$. In fact, if we make the replacement $\int \frac{ d\Omega}{2\pi} \rightarrow \sum_m, \, \Omega \rightarrow \Omega_m = 2 \pi m T$ in Eq. (\ref{magfield}), we can directly calculate the contribution at finite temperature and finite magnetic field. Doing this, we find
\begin{align}
\nonumber \delta \chi_{dia}^{q=0} (H,T) = \chi_{dia}^{0}\frac{U^2 N_F^2}{8} \frac{\mu_B H} { E_F}  \csch^2  \left( \frac{\mu_B H}{T} \right) \\
\times \left[ \sinh \left( 2\frac{\mu_B H}{T} \right) - \frac{\mu_B H}{T}\right].
\label{ss_11}
\end{align}
 To find the total scaling function, we must also take into account the contributions from $q=2k_F$. We recall that there is a nonanalytic contribution from $q=2k_F$ at finite temperature, but not at finite Zeeman field. Taking this into account, we obtain the total
\begin{align}
\nonumber \delta \chi_{dia} (H,T) = \chi_{dia}^{0}\frac{U^2 N_F^2}{8} \frac{T} { E_F} \Bigg(  \frac{\mu_B H}{T} \csch^2  \left( \frac{\mu_B H}{T} \right) \\
\times \left[ \sinh \left( 2\frac{\mu_B H}{T} \right) - \frac{\mu_B H}{T}\right] +1 \Bigg).
\end{align}

We note that the $q=0$ contribution to the diamagnetic susceptibility contains the same scaling function as the paramagnetic susceptibility. The overall scaling function is different due to the presence of the $q=2k_F$ contribution in the diamagnetic susceptibility.
Adding both the paramagnetic and diamagnetic contributions together into the total magnetic susceptibility, we find
\begin{align}
\nonumber \chi (H,T) = \frac{2}{3}& \chi^{0}_{para} \left(1 + \frac{3}{2} N_F U + 1.09 (N_F U)^2\right) \\
&+ \frac{11}{12} \delta \chi_{para}(H,T) - \frac{1}{12} \delta \chi_{para}(0,T),
\label{u_4}
\end{align}
where $\delta \chi_{para} (H,T)$ is given by (\ref{ss_10}).

We make a few remarks on the extension of these calculations to 3D. In the case of the spin susceptibility, there has found to be a $H^2 \log H$ contribution to the spin susceptibility \cite{betouras2005thermodynamics}. Since the calculations to the diamagnetic susceptibility have paralleled the spin susceptibility in the 2D case, we would then expect there to be an analogous nonanalyticity in the diamagnetic susceptibility. The calculations in for finite Zeeman field in 3D for the diamagnetic susceptibility would proceed in the same ways as we have done above, i.e. assuming that momentum transfers are close to $q=0$ or $q=2k_F$, and evaluating the relevant diagrams in these limits. In the case of finite temperature, one needs to be more careful. For the spin susceptibility, one would expect a $T^2 \log T$ term analogous to the $H^2 \log H$ term. However, this is not the case - in 3D, there is no $T^2 \log T$ nonanalyticity in the spin susceptibility. One is not able to determine whether this is the case for the diamagnetic susceptibility immediately based on our results. Further calculations in 3D are necessary in order to determine if such a $T^2\log T$ term could be present in the diamagnetic susceptibility.

\section{Conclusions}
We have analyzed the Landau diamagnetic susceptibility diagrammatically for a model of 2D fermions with  Hubbard-like interaction. We used the relation $\chi_{dia}  = (e/c)^2 \lim_{Q\rightarrow0} \Pi^{JJ}_\perp (Q)/Q^2$, where
$\Pi^{JJ}_\perp$ is the transverse component of the static current-current correlator. For free fermions, we reproduced diagrammatically the Landau-Peierls formula for arbitrary fermionic dispersion (it reduces to
  $\chi_{dia} = - (2\mu^2_B) N_F/3$ for a parabolic dispersion).
   For interacting fermions, we evaluated
$\Pi^{JJ}_\perp (Q)$ up to second order in Hubbard $U$ by combining
self energy, Maki-Thompson, and Aslamazov-Larkin-type diagrams. At first order in $U$, we found no correction to the diamagnetic susceptibility. At order $U^2$, we obtained a regular correction $\delta \chi_{dia} \propto U^2$ at zero temperature and zero magnetic field, and explicitly obtained the prefactor.   In the process of calculations, we found that individual diagrams for $\Pi^{JJ}_\perp (Q)$ contain non-analytic $|Q|$ terms, but these terms cancel out in the full expression, and $\Pi^{JJ}_\perp (Q) \propto Q^2$.  In this  respect, $\Pi^{JJ}_\perp (Q)$ behaves similarly to charge polarization, for which $|Q|$ terms from individual diagrams also cancel out.

  We next considered the corrections to the prefactor of the $U^2$ term in $\delta \chi_{dia}$ in both temperature and magnetic field. We showed that the Landau diamagnetic susceptibility does indeed have nonanalytic linear in $T$ and linear in $H$ terms. In this respect, the behavior of the diamagnetic susceptibility is similar to that of the paramagnetic Pauli susceptibility, which also contains such terms.  We computed analytically the prefactors for $O(U^2 T)$ and $O(U^2 H)$  terms in $\delta \chi_{dia}$ for parabolic fermionic dispersion.
   We found that for both finite temperature and magnetic field, the nonanalytic contributions are of the same sign as the bare Landau diamagnetic susceptibility, and therefore serve to enhance the diamagnetic effects as temperature and magnetic field increase. By magnitude, nonanalytic corrections to the diamagnetic susceptibility are comparable  to non-analytic corrections to the spin susceptibility.

\section{Acknowledgments}
We thank D.L. Maslov for useful discussions and comments. We also thank Keiya Shirahama for bringing our attention to references in Ref. \footnotemark[1].
The research  was supported by the U.S. Department of Energy, Office of
Science, Basic Energy Sciences, under Award No.\ DE-SC0014402.

\onecolumngrid
\appendix

\section{Reproducing the Landau-Peierls Formula from Diagrammatics}
\label{landaupeierls}
In this Appendix we derive diagrammatically the Landau-Peierls formula for diamagnetic susceptibility of free fermions with arbitrary dispersion. The calculation  has been performed in collaboration with D. L. Maslov.

The formula for the diamagnetic susceptibility in a free electron gas has been derived by Landau in 1930 \cite{landau2013statistical}.
Three years later, Peierls obtained a correction to this expression when the electrons experience a periodic potential due to ions  \cite{peierls, wilson1936metals}. This correction is usually written as the sum of the contribution from the ions and from conduction electrons. The contribution from the conduction electrons is
\begin{align}
\chi_{dia} = \frac{2 \mu_B^2 m^2}{3(2 \pi)^d} \int d^d k \, n_F'(\e_k) \left( \frac{\partial^2 \e_{\vb k }}{\partial k_x^2}  \frac{\partial^2 \e_{\vb k }}{\partial k_y^2} - \left( \frac{ \partial^2 \e_{\vb k}}{\partial k_x \partial k_y}\right)^2\right),
\label{v_5}
\end{align}
where $n_F'(\e_k) = \frac{\partial n_F}{\partial \e_k}$ and $d=2,3$. We note that as $T\rightarrow0$, $n_F'(\e_{\vb k}) \rightarrow - \delta(\e_{\vb k})$, so at sufficiently small $T$, the above integral is equivalent to averaging the integrand over the Fermi surface.  Eq. (\ref{v_5}) is known as Landau-Peierls formula.

It is well known that the Landau diamagnetic susceptibility  $\chi_{dia}$  for free fermions with a parabolic dispersion can be reproduced diagrammatically by expressing $\chi_{dia}$ via the transverse component of the static current-current correlator, $\Pi^{JJ}_\perp (Q)$ as $\chi_{dia}  = (e/c)^2 \lim_{Q\rightarrow0} \Pi^{JJ}_\perp (Q)/Q^2$ (Eq. (\ref{dia}) in the main text), and evaluating $\Pi^{JJ}_\perp (Q)$ as the particle-hole bubble with transverse velocity
 $v_{\vb k}^{\perp}$ in the vertices \cite{pinesnozieres}.
Our aim here is to show that the diagrammatic formalism can also reproduce Eq. (\ref{v_5}) for an arbitrary dispersion relation.

For arbitrary dispersion, the particle-hole current-current bubble is given by
\begin{align}
\Pi^{JJ}_\perp(Q \hat x,0) &= -2 T \sum \limits_{\omega_n} \int \frac{d^dk}{(2\pi)^d} \left( v_{\vb{k}}^y \right)^2 G (\vb{k} + \vb Q/2, \omega_n)G( \vb k - \vb Q/2,\omega_n) - \cdots \\
\label{free}&= -2 \int \frac{d^dk}{(2\pi)^d} \left( v_{\vb{k}}^y \right)^2 \frac{n_F(\e_{\vb{k} + \vb Q/2 }) - n_F(\e_{\vb{k}-\vb Q/2})}{\e_{\vb k+\vb Q/2}-\e_{\vb{k} - \vb Q/2}} - \cdots,
\end{align}
where dots stand for the $Q=0$ terms that needs to be subtracted.
To get the $Q^2$ term in $\Pi^{JJ}_\perp(Q \hat x,0)$, we must expand each term in the r.h.s of (\ref{free}) to order $Q^2$. Doing so, we find
\begin{align}
v_k^y &= \frac{\partial \e_{\vb k}}{\partial k_y} = \partial_y \e_{\vb k}\\
\e_{\vb k \pm \vb Q/2} &= \e_{\vb k} \pm \frac{Q}{2} \partial_x \e_{\vb k} + \frac{Q^2}{8} \partial_x^2 \e_{\vb k} \pm \frac{Q^3}{48} \partial_x^3 \e_{\vb k}\\
\e_{\vb k + \vb Q/2} - \e_{\vb k - \vb Q/2} &= Q \partial_x \e_{\vb k} + \frac{Q^3}{24} \partial_x^3 \e_{\vb k}\\
\frac{n_F(\e_{\vb k +\vb Q/2}) - n_F(\e_{\vb k - \vb Q/2})}{\e_{\vb k + \vb Q/2} - \e_{\vb k - \vb Q/2}} &= n_F'(\e_{\vb k} ) + \frac{Q^2}{24} \left(3 n_F''(\e_{\vb k}) \partial_x^2 \e_{\vb k} + n_F'''(\e_{\vb k}) \left( \partial_x \e_{\vb k} \right)^2 \right),
\end{align}
where have made use of the notation $\frac{\partial}{\partial k_i} = \partial_i$. Inserting these expressions into Eq. (\ref{free}), subtracting off the $Q^0$ term, and using Eq. (\ref{dia}) to find the diamagnetic susceptibility, we have
\begin{align}
\chi_{dia} = -\frac{\mu_B^2 m^2}{3(2 \pi)^d} \int d^dk (\partial_y \e_{\vb k})^2 \left( 3 n_F''(\e_{\vb k}) \partial_x^2 \e_{\vb k} + n_F'''(\e_{\vb k}) (\partial_x \e_{\vb k})^2\right)
\end{align}
Examining first the term proportional to $n_F'''(\e_{\vb k})$, the expression can be simplified by using the chain rule to write $n_F'''(\e_{\vb k}) \partial_x \e_{\vb k} = \partial_x n_F''(\e_{\vb k})$ and then integrating by parts to find
\begin{align}
\int d^dk \, n_F'''(\e_{\vb k})  (\partial_y \e_{\vb k})^2 (\partial_x \e_{\vb k})^2 = - \int d^dk \, n_F''(\e_{\vb k})  \frac{\partial}{\partial k_x} \left( (\partial_y \e_{\vb k})^2 (\partial_x \e_{\vb k}) \right)
\end{align}
Simplifying the above expression and inserting it back into the diamagnetic susceptibility, we have
\begin{align}
\chi_{dia} = -\frac{\mu_B^2 m^2}{6 \pi^2} \int d^dk\, n_F''(\e_{\vb k})  \left( \partial_x^2 \e_{\vb k}  \left(\partial_y \e_{\vb k} \right)^2 - \partial_x \e_{\vb k} \partial_y \e_{\vb k} \,  \left( \partial^2_{xy} \e_{\vb k} \right) \right)
\end{align}
Again using chain rule to write $n_F''(\e_k) \partial_y \e_{\vb k} = \partial_y n_F'(\e_{\vb k})$ for the first term and $n_F''(\e_k) \partial_x \e_{\vb k} = \partial_x n_F'(\e_{\vb k})$ for the second term, then integrating by parts once more, we get
\begin{align}
\chi_{dia} = \frac{2\mu_B^2 m^2}{3(2 \pi)^d } \int d^dk \, n_F'(\e_k) \left( \frac{\partial^2 \e_{\vb k }}{\partial k_x^2}  \frac{\partial^2 \e_{\vb k }}{\partial k_y^2} - \left( \frac{ \partial^2 \e_{\vb k}}{\partial k_x \partial k_y}\right)^2\right).
\label{aaa}
\end{align}
This is precisely the Landau-Peierls formula for the conduction electron part of the diamagnetic susceptibility, Eq. (\ref{v_5}).
If we take the ratio of $\chi_{dia}$ and $\chi_{para}$ for free fermions with an arbitrary dispersion relation, we find
\begin{align}
\frac{\chi_{dia}}{\chi_{para}} = -\frac{1}{3} \frac{m^2}{\int d^dk\, n_F'(\e_k)} \int d^dk \, n_F'(\e_k) \left( \frac{\partial^2 \e_{\vb k }}{\partial k_x^2}  \frac{\partial^2 \e_{\vb k }}{\partial k_y^2} - \left( \frac{ \partial^2 \e_{\vb k}}{\partial k_x \partial k_y}\right)^2\right).
\end{align}
The factor of $-\frac{1}{3}$ emerges when the quantity $ \frac{\partial^2 \e_{\vb k }}{\partial k_x^2}  \frac{\partial^2 \e_{\vb k }}{\partial k_y^2} - \left( \frac{ \partial^2 \e_{\vb k}}{\partial k_x \partial k_y}\right)^2$ is $m^{-2}$, that is when the dispersion is parabolic.

We emphasize that Eq. (\ref{aaa}) is for a single band of fermions.  In materials with multiple bands,  such as bismuth and graphene, the Landau-Peierls formula does not give the full contribution to the diamagnetic susceptibility~\cite{adams1953magnetic, mcclure1956diamagnetism}. Besides, when electrons are in a periodic potential, there is an additional term expandable in powers of the density that is not included in the Landau-Peierls formula~\cite{kjeldaas1957theory,cornean2011rigorous}. This term  describes Langevin diamagnetism in the limit of tight-binding and is relevant when the density of fermions is not small.
Still, there is evidence from ab initio calculations that, in at least some materials like the alkali metals, the Landau diamagnetic susceptibility is adequately described by the Landau-Peierls formula.
~\cite{nikolaev2018landau}.

\section{$|Q|$ Nonanalyticities of Self Energy and Maki-Thompson Diagrams}
\label{nonanalyticity}
Here we explicitly calculate $|Q|$ nonanalyticities in the current-current correlator for the diagrams (a) and (c) in Fig. \ref{secondorder}, i.e. the second order self energy and Maki-Thompson diagrams. From similar calculations for the charge and spin susceptibilities, we know that these nonanalyticities comes from contributions of small momentum and frequency transfers as well as momentum transfers close to $2k_F$. We begin by considering these small momentum transfers in both the case of the second order self energy and Maki-Thompson diagrams, then consider the backscattering case in these diagrams afterwards.
To simplify notations, below we re-label $\Pi^{JJ}_{\perp}$ by just $\Pi$.
\subsection{Nonanalyticity from small $q$ }
We first consider the case of the self energy diagram, which is given by
\begin{align}
\Pi_{a}(Q,0) = 8 U^2 \int_{k,q} \Pi(q,\Omega_m) \left(v_k^y\right)^2 G_{k+Q/2}^2 G_{k-Q/2} G_{k+q+Q/2}
\end{align}
For small $q$ and $Q$, all contributions come from close to the Fermi surface, so we can make the approximations
\begin{align}
\e_{k} &= v_F (k-k_F)\\
\e_{k \pm Q/2} &= \e_{k} \pm \vb {v}_F \cdot \vb Q/2\\
\e_{k + q \pm Q/2} &= \e_{k} + \vb {v}_F \cdot \vb q \pm \vb {v}_F \cdot \vb Q/2\\
v_k^y &= v_F \sin \theta
 \end{align}
where $\vb{v}_F = v_F \hat k$, and $\theta = \angle(\vb k, \vb Q)$. Noting the expression is even over $\Omega_m$ so that we can reduce the expression to an integral over $\Omega_m$ from $0$ to $\infty$, and then integrate first over $\e_k$. The integrals over $\omega_n$ and $\theta$ are also elementary, and can be evaluated to give
\begin{align}
\label{firstint} \delta \Pi^{q=0}_{a}(Q,0) = \frac{U^2 v_F^2 m }{\pi^4} \int d^2q \int_{0}^{\infty} d \Omega_m \,  \Pi(\vb{q}, \Omega_m) \frac{i \Omega_m}{v_F^2 Q^2  (i\Omega_m - \vb {v}_F \cdot \vb q )^2}\\
\nonumber \times \left( i \Omega_m -  \vb {v}_F \cdot \vb q - i \sqrt{v_F^2 Q^2 - (i \Omega_m - \vb {v}_F \cdot \vb q )^2} \right),
\end{align}
where we note that in the small $q$ approximation, we can write
\begin{align}
\Pi(q, \Omega_m) = \frac{m}{2\pi} \left( -1 + \frac{\Omega_m}{\sqrt{ v_F^2 q^2 + \Omega_m^2}}\right).
\end{align}
We make the change to polar coordinates here, so that $\Omega_m = r \sin \phi$ and $q = r \cos \phi$. We also rescale $r$ to be in units of $Q$, so that the total function is now
\begin{align}
 \frac{i U^2 k_F |Q|  }{\pi^4}\int_{0}^{\pi/2} d\phi \int_{0}^{2\pi}d \xi & \int_{0}^{\infty} dr \Pi(\phi) \sin \phi \cos \phi \frac{r}{ (i \sin \phi - \cos \phi \cos \xi)^2} \\
 \nonumber &\times \left(r (i \sin \phi - \cos \phi \cos \xi) - i \sqrt{1 -  r^2 (i \sin \phi - \cos \phi \cos \xi )^2} \right),
\end{align}
where $\xi = \angle(\vb k, \vb q)$ and $\Pi(\phi) = m/2\pi \left(-1 + \sin \phi \right)$. We note that this term is divergent over $r$, and needs a cutoff $r_{max}$. However, we are only interested in the nonanalytic contribution of this term, which is a low energy contribution independent of the cutoff. Integrating over $r$ and negelecting the cutoff dependent terms leaves only
\begin{align}
\nonumber \delta \Pi^{q=0}_{a}(Q,0) = &-\frac{U^2 k_F |Q| }{3\pi^4}\int \limits_0^{\pi/2} d\phi \,\int \limits_0^{2\pi} d \xi \Pi(\phi) \sin \phi \cos \phi \frac{1}{(i \sin \phi - \cos \phi \cos \xi)^4}\\
\nonumber = &-\frac{U^2 k_F m |Q| }{24\pi^4}\int \limits_0^{\pi/2} d\phi (-1+\sin \phi) \sin \phi \cos \phi \left(5 \sin 3 \phi - 3 \sin \phi \right) \\
= &\frac{U^2 k_F m }{72 \pi^4} |Q|.
\end{align}
Now we can also consider the Maki-Thompson contribution. The diagram gives
\begin{align}
\Pi_{b}(Q,0) = 4 U^2 \int_{k,q} \Pi(q,\Omega_m) v_k^y v_{k+q}^y G_{k+Q/2} G_{k-Q/2} G_{k+q+Q/2} G_{k+q-Q/2}
\end{align}
We note that to lowest order, we have $v_{k+q}^y = v_{k}^y$, as the additional corrections due to $q$ yield only contributions to orders $Q^2$ and higher, so we may discard them. Knowing this, we can integrate over $\e_k$, $\omega_n$, and $\theta$ to give
\begin{align}
-\frac{U^2 v_F^2 m }{\pi^4}\int \; d^2q \int_{0}^{\infty} d\Omega_m &\Pi(\vb{q}, \Omega_m) \frac{i \Omega_m}{v_F^2 Q^2  (i\Omega_m - \vb {v}_F \cdot \vb q )^2} \\
\nonumber &\times \left( i \Omega_m -  \vb {v}_F \cdot \vb q - i \sqrt{v_F^2 Q^2 - (i \Omega_m - \vb {v}_F \cdot \vb q )^2} \right)
\end{align}
Comparing the above equation to Eq. (\ref{firstint}), we find that this contribution exactly cancels with the contribution from the self energy term. Therefore, there is no net nonanalyticity at $q\sim0$ for these two diagrams.
\subsection{Nonanalyticity from momenta near $2k_F$ }
We now consider the nonanalyticities which come from
 momentum  transfers
 close to $2k_F$, corresponding to backscattering. In this case, we can approximate $\e_{k+q} = -\e_k + v_F \tilde q + 2 v_F k_F (1+ \cos \theta)$, where $ \tilde q = (q -2k_F)$  and  $\hat k \cdot \hat q = \cos \theta$. In addition, the dominant contributions will come from angles close to perfect backscattering, so we can additionally write $ 1+\cos \theta = \frac{1}{2}(\pi - \theta )^2 = \frac{1}{2} \tilde \theta^2$. Then, we can write the self energy contribution as
\begin{align}
\nonumber \delta \Pi^{2k_F}_{a}(Q,0) &=8 U^2 \int_{k,q} (v_k^y)^2\Pi(q,\Omega_m) \left(G_{k+Q/2} \right)^2 G_{k-Q/2} G_{k+q+Q/2} \\
&= 8 U^2 \int_{k,q} \Pi(q, \Omega_m) \left(v_F \sin \theta_1 \right)^2  \left( \frac{1}{i \omega_n - \e_{k+Q/2} } \right)^2  \\
\nonumber &\times \frac{1}{i \omega_n - \e_{k- Q/2} } \frac{1}{i (\omega_n + \Omega_m)+ \e_{k+Q/2} - v_F \tilde q - v_F k_F\tilde \theta^2 },
\end{align}
We can rescale $\e_k$, $\omega_n$, $\Omega_m$, $\tilde q$, and $\tilde \theta$ to be unitless, and then integrate over $\e_k$. Doing so, after some simplification, we find
\begin{align}
\nonumber \frac{2 U^2 k_F |Q| m }{\pi^6}\int \limits_{-\infty}^{\infty} d \tilde q \int \limits_{0}^{\infty} d \Omega_m \int \limits_{0}^{\pi} d \theta_1 \int \limits_{0}^{\infty} d \tilde \theta \left( \sqrt{ \tilde q +i \Omega_m} + \sqrt{ \tilde q - i \Omega_m} \right) \sin^2 \theta_1 \\
\text{Im} \left( \int \limits_{0}^{\infty}d\omega_n \frac{1}{\left(i(\omega_n+\Omega_m) - \tilde q - \tilde \theta^2\right)^2 \left( i(\omega_n + \Omega_m) - \tilde q - \tilde \theta^2 + \cos \theta_1\right)} \right),
\end{align}
where we have written $\Pi(q,\Omega_m)$ as $m/2\pi \left( \sqrt{ \tilde q +i \Omega_m} + \sqrt{ \tilde q - i \Omega_m}  \right) $. We can then convert this to polar coordinates, with $\tilde q = r \cos \phi$ and $\Omega_m = r \sin \phi$. Rescaling variables so that $ \omega_n \rightarrow r \omega_n$ and $ \tilde \theta \rightarrow \sqrt{r} \tilde \theta$, we get
\begin{align}
\frac{2U^2 k_F |Q| m }{\pi^6} \text{Im}\int \limits_{0}^{\pi} d\phi \int \limits_{0}^{\pi} \theta_1 \int \limits_{0}^{\infty} d \tilde \theta \int \limits_{0}^{\infty} d\omega_n & \int \limits_{0}^{\infty} dr \cos \phi/2 \sin^2 \theta_1\\
\nonumber \times &\frac{1}{\left(i \omega_n - e^{-i \phi} - \tilde \theta^2 \right)^2} \frac{r}{r\left(i \omega_n - e^{-i \phi} - \tilde \theta^2 \right) + \cos \theta_1 }
\end{align}
Integrating over both $r$, taking only the low energy, cutoff independent term, and then integrating over $\theta_1$, we find
\begin{align}
\label{2kfselfenergy} - \frac{2U^2 k_F |Q| m }{3\pi^5} \text{Re}\int \limits_{0}^{\pi} d \phi \int \limits_{0}^{\infty} d \tilde \theta \int \limits_{0}^{\infty} d\omega_n \cos \phi/2 \frac{1}{\left(i \omega_n - e^{-i \phi} - \tilde \theta^2 \right)^4}
\end{align}
The remaining integrals are elementary, and give the result
\begin{align}
\delta \Pi^{2k_F}_a(Q,0) = \frac{k_F m U^2}{72 \pi^4} |Q|.
\end{align}
Now, for the second order Maki-Thompson result. We can write out this contribution as
\begin{align}
\nonumber \delta \Pi_{b}^{2 k_F}(Q,0) & = 4 U^2 \int_{k,q} \Pi(q,\Omega_m) v_k^y v_{k+q}^y G_{k+Q/2} G_{k-Q/2} G_{k+q+Q/2} G_{k+q-Q/2} \\
&= - 4 U^2 \int_{k,q} \Pi(q, \Omega_m) \left(v_F \sin \theta_1 \right)^2  \frac{1}{i \omega_n - \e_{k+Q/2} } \frac{1}{i \omega_n - \e_{k- Q/2} } \\
\nonumber & \times \frac{1}{i (\omega_n + \Omega_m)+ \e_{k+Q/2} - v_F \tilde q - v_F k_F\tilde \theta^2 } \frac{1}{i (\omega_n + \Omega_m)+ \e_{k-Q/2} - v_F \tilde q - v_F k_F\tilde \theta^2 }
\end{align}
We note here that, like in the case of the $q \sim 0$ nonanalyticity, we can neglect the contribution of $\tilde q$ to $v_{k+q}$. In addition, we can take the direction of $\vb{q}$ to be exactly antiparallel to $\vb{k}$, as including deviations from $\vb{q} = -\vb{k}$ will also only contribute at higher orders of $|Q|$. However, for $q \sim 2k_F$, $\e_{k+q} =  -\e_k + v_F \tilde q + 2 v_F k_F (1+ \cos \theta)$, so neglecting contributions of order $\tilde q$ and setting $\theta = \pi$ means $\e_{k+q} = -\e_k$. Then, $v_k^y v_{k+q}^y =-(v_k^y)^2$. We rescale variables as before, and then integrate over $\e_k$. Then, converting to polar coordinates, we find
\begin{align}
\nonumber \frac{U^2 k_F |Q| m }{2 \pi^6}  \text{Im} \int_{0}^{\phi} d\phi \int_{0}^{\pi} d \theta_1 \int_{0}^{\infty}d \tilde \theta \int_{0}^{\infty} d\omega_n \int_{0}^{\infty} &dr \cos \phi/2 \sin^2 \theta_1\frac{1}{\left(i \omega_n - e^{-i \phi} - \tilde \theta^2 \right)^2} \\
 \times &\frac{r^2}{r^2\left(i \omega_n - e^{-i \phi} - \tilde \theta^2 \right)^2 - \cos^2 \theta_1}
\end{align}
Now, we can integrate over $r$ and $\theta_1$ as before, discarding the cutoff dependent term, getting
\begin{align}
\frac{2U^2 k_F |Q| m }{3\pi^5} \text{Re}\int \limits_{0}^{\pi} d \phi \int \limits_{0}^{\infty} d \tilde \theta \int \limits_{0}^{\infty} d\omega_n \cos \phi/2 \frac{1}{\left(i \omega_n - e^{-i \phi} - \tilde \theta^2 \right)^4}
\end{align}
Comparing with Eq. (\ref{2kfselfenergy}), this is precisely the same contribution as the self energy term with a minus sign. Therefore, the nonanalyticity in the Maki-Thompson diagram from momentum transfers of $2k_F$ is
\begin{align}
\delta \Pi_{b}^{2 k_F}(Q,0)=-\frac{k_F m U^2}{72 \pi^4} |Q|.
\end{align}
With this results, we have confirmed there is no net nonanalyticity between the self energy and Maki-Thompson diagrams for both $q=0$ and $q =2k_F$.

\section{Sum of Diagrams (c) and (g)}
\label{ap1}
In this appendix, we consider the sum of diagrams (c) and (g) at finite temperature and magnetic field. We show that at order $Q^2$, the two diagrams in fact cancel. At zero magnetic field, diagram (c) and diagram (f) are equal, so this calculation also confirms that when $H=0$, the pair of Aslamazov-Larking diagrams cancel.

We can write the sum of these two diagrams as
\begin{align}
\label{alsum} \Pi_{c}+\Pi_g =2 U^2 T \sum_{\alpha \neq \beta} \sum_{\Omega_m} \int_{\vb q} I_{\alpha \beta}(\vb Q, \vb q,\Omega_m) \left(I_{\alpha \beta}(\vb Q, \vb q,\Omega_m) + I_{\beta \alpha}(\vb Q, - \vb q,-\Omega_m) \right),
\end{align}
where $\int_{\vb q} = \int d^2q/(2\pi)^2$ and $ I_{\alpha \beta}(\vb Q, \vb q,\Omega)$ is a triad of Green's functions defined as
\begin{align}
\label{triad} I_{\alpha \beta}(\vb Q, \vb q,\Omega_m) = \sum_{\omega_n} \int_{\vb k} v_k^y G^{\alpha}_{k+Q/2} G^{\alpha}_{k-Q/2} G^{\beta}_{k+q}.
\end{align}
Symmetrizing Eq. (\ref{alsum}) with respect to $\Omega_m$, and also noting we can exchange $\alpha$ and $\beta$ in the sum, we can rewrite the expression as
\begin{align}
U^2 T \sum_{\alpha \neq \beta} \sum_{m=1}^{\infty}\int_{\vb q}  \left(I_{\alpha \beta}(\vb Q, \vb q,\Omega_m) + I_{\beta \alpha}(\vb Q, - \vb q,-\Omega_m) \right)^2.
\label{w_1}
\end{align}
We have left out the $\Omega_m=0$ term in this sum. One can confirm that this term is zero by following the same steps that we outline below to show all $\Omega_m \neq 0$ terms are zero. Series expanding
 Eq. (\ref{w_1}),
 we can see the term proportional to $Q^2$ has the form
\begin{align}
\frac{1}{2} U^2 Q^2 T \sum_{\alpha \neq \beta} \sum_{m}\int_{\vb q} \left(I_{\alpha \beta} (0,\vb q,\Omega_m) + I_{\beta \alpha}(0,-\vb q,-\Omega_m) \right)\left(I_{\alpha \beta}''(0,\vb q,\Omega_m) + I_{ \beta \alpha}''(0,-\vb q,-\Omega_m) \right),
\end{align}
where $I_{\alpha \beta}''(0,\vb q,\Omega_m) = \lim_{Q\rightarrow0} \frac{d^2}{d Q^2} I_{\alpha \beta}(\vb Q,\vb q,\Omega_m)$, and we have dropped terms proportional to $I_{\alpha \beta}'(0,\vb q,\Omega_m)$ as these go to zero. We claim that $I_{\alpha \beta}(0,\vb q,\Omega_m) + I_{\beta \alpha}(0,-\vb q,-\Omega_m) =0,$ so that the entire $Q^2$ term vanishes. To show this, we sum over over fermionic Matsubara frequencies, and then take the limit as $Q\rightarrow 0$. Doing so, we find
\begin{align}
\nonumber I_{\alpha \beta}(0,&\vb q,\Omega_m) + I_{\beta \alpha}(0,-\vb q,-\Omega_m) = \\
&\int_{\vb k}  \left(\frac{v_k^y \left(n_F \left(\e _{k+q}^{\beta}\right)-n_F \left(\e _k^{\alpha}\right)\right)}{\left(i \Omega_m-\e _{k+q}^{\beta}+\e_k^{\alpha}\right){}^2}+\frac{v_k^y n_F' \left(\e _k^{\alpha} \right)}{i \Omega_m -\e _{k+q}^{\beta}+\e _k^{\alpha}} - (c.c., \alpha \leftrightarrow \beta)\right),
\end{align}
where second term denotes the complex conjugate of the first term with $\alpha$ and $\beta$ interchanged, and we have used the fact that $I_{\beta \alpha}(0,-\vb q,-\Omega_m) = -I_{\beta \alpha}(0,\vb q,-\Omega_m)$. One can derive this relation from Eq. (\ref{triad}) by making the transformation $\vb k \rightarrow -\vb k$. To evaluate the terms proportional to $n_F \left(\e_{k+q}^{\alpha (\beta)} \right)$, we can make the transformation $\vb k\rightarrow -\vb k -\vb q$ so that $n_F \left(\e_{k+q}^{\alpha (\beta)} \right) \rightarrow n_F \left(\e_{k}^{\alpha (\beta)} \right)$. Doing so and simplifying the expression, we find
\begin{align}
\frac{1}{m} \int_{\vb k} \left(\frac{q_y n_F \left(\e _k^{\alpha}\right)}{\left(i \Omega_m-\e _{k+q}^{\beta}+\e _k^{\alpha} \right){}^2}+\frac{k_y n_F' \left(\e _k^{\alpha}\right)}{i \Omega_m-\e _{k+q}^{\beta}+\e_k^{\alpha} } - (c.c., \alpha \leftrightarrow \beta) \right)
\end{align}
We can explicitly write out $q_y = q \sin \theta_{qQ}$, $ \e _k-\e _{k+q} = - \frac{1}{m}k q \cos \theta_{kq} - \frac{q^2}{2m} \pm 2 \mu_B H$, and $k_y  = k \sin (\theta_{qQ} + \theta_{kq})$. The sign of $\mu_B H$
 determines
 whether $\alpha = \uparrow, \beta = \downarrow$ or $\alpha = \downarrow, \beta = \uparrow$. Then, integrating over $\theta_{kq}$, we find
\begin{align}
\label{aldone}  \int \frac{dkk}{2\pi} \frac{ m \sin \theta_{qQ}}{q} \left(\frac{i \xi n_F(\e_k^{\alpha})}{\left(k^2-\xi^2 \right)^{3/2}}  -\frac{1}{m}\left(1 + \frac{i\xi}{\sqrt{k^2-\xi^2}} \right)n_F'(\e_k^{\alpha}) - (c.c., \alpha \leftrightarrow \beta)\right),
\end{align}
where we have written $\xi = \frac{i \Omega_m \pm \mu_B H }{q} -\frac{q}{2}$. Formally, the above expression also depends on the sign of the imaginary part of $\xi$. However, since we initially symmetrized the function so that $\Omega_m>0$, we can simply write the function as it is shown above. Lastly, we can do integration by parts on the term proportional to $n_F'(\e_k)$ to combine both above terms. Doing so, we find
\begin{align}
-\int dkk \frac{1}{m}\left(1 + \frac{i \xi}{\sqrt{k^2-\xi^2}} \right)n_F'(\e_k^{\alpha})  &= -\frac{1}{m}\left(1+ \frac{i \xi}{\sqrt{k^2 - \xi^2}}\right) n_F(\e_k^{\alpha}) \bigg \rvert_{k = 0}^{k=\infty} \\
\nonumber &- \int dk k \frac{i\xi n_F(\e_k^{\alpha})}{\left(k^2 - \xi^2 \right)^{3/2}},
\end{align}
We can see this second term exactly cancels the term proportional to $n_F(\e_k)$ in Eq.
 (\ref{aldone}), leaving only the term at the bounds,
\begin{align}
-\left(1+ \frac{i\xi}{\sqrt{-\xi^2}}\right) n_F(\e_{k=0}^{\alpha}).
\end{align}
However, we can simplify this term further yet. Recalling that $\Omega_m>0$, and noting the branch cut that occurs in the square root, we can rewrite $\frac{i\xi}{\sqrt{-\xi^2}} = -1$ so that the entire expression is zero. Therefore, we only need diagrams (a) and (f) when calculating the $Q^2$ term of the current-current correlation function, even when both temperature and external magnetic field are finite.

\section{Evaluation of $\delta\chi_{dia} (H)$}
\label{magfield}
The point of departure is Eq. (\ref{finiteT1}) in the main text:
\begin{align}
 \delta \Pi^{JJ}_\perp(H)=U^2 \sum_{\alpha \neq \beta} \int_{ k,  q}  \Big[\Pi^{\alpha \beta} (q,\Omega_m) \Big( 2 (v_k^y)^2 \left(G^{\alpha}_{k+Q/2}\right)^2 G^{\alpha}_{k-Q/2} G^{\beta}_{k+q+Q/2}\nonumber \\
 + v_k^y v_{k+q}^y G^{\alpha}_{k+Q/2} G^{\alpha}_{k-Q/2} G^{\beta}_{k+q+Q/2}G^{\beta}_{k+q-Q/2}\Big)\Big]
\end{align}
We re-express it as
\begin{align}
\nonumber \delta \Pi^{JJ}_\perp(H)= & U^2 \sum_{\alpha \neq \beta} \int_{ k,  q} \Big[\Pi^{\alpha \beta} (q,\Omega_m) \Big( (v_k^y)^2 G^{\alpha}_{k+Q/2}  G^{\alpha}_{k-Q/2} G^{\beta}_{k+q+Q/2} \left(2 G^{\alpha}_{k+Q/2} +G^{\beta}_{k+q-Q/2}\right) \\
\nonumber & \hspace{140pt}+  v_k^y v_{q}^y G^{\alpha}_{k+Q/2} G^{\alpha}_{k-Q/2} G^{\beta}_{k+q+Q/2}G^{\beta}_{k+q-Q/2}\Big)\Big] \\
= &\frac{U^2}{2} \sum_{\alpha \neq \beta} \int_{k, q}  \Pi^{\alpha \beta} (q,\Omega_m) \left( \tilde G_1 + \tilde G_2 \right),
\end{align}
where we have used the fact that $v_{k+q}^y = (k_y + q_y)/m = v_k^y + v_q^y$ for a parabolic dispersion. Since we are only interested in the nonanalytic contribution to function when $q=0$, we can easily integrate over $\e_k$ first. Therefore, unlike in the case of $H=0, T=0$, we can series expand in $Q$ before integrating as long as the integral over $\e_k$ is done before the integral over frequency. We can therefore expand $\tilde G_1$ and $\tilde G_2$ to order $Q^2$ in a manner that is similar calculation of the gradient term of the spin and charge susceptibilities \cite{maslov2017gradient}. Expanding $\tilde G_1$, we find a total of four terms, such that $\tilde G_1 = \tilde G_1^a+ \tilde G_1^b+\tilde G_1^c+\tilde G_1^d$, where
\begin{align}
\tilde G_1^{a} &= \frac{1}{2m^2} \left(v_k^y \right)^2 \left( \vb k \cdot \vb Q\right)^2 \left( G_{k,\alpha}^2 G_{k+q,\beta}^4 + 2 G_{k,\alpha}^3 G_{k+q,\beta}^3 + 3 G_{k,\alpha}^4 G_{k+q,\beta}^2 +4 G_{k,\alpha}^5 G_{k+q,\beta} \right)\\
\tilde G_1^{b} &= \frac{Q^2}{2m} \left(v_k^y \right)^2  \left(G_{k,\alpha}^2 G_{k+q,\beta}^3 + 2 G_{k,\alpha}^3 G_{k+q,\beta}^2 + 3 G_{k,\alpha}^4 G_{k+q,\beta} \right)\\
\tilde G_1^{c} &= \frac{1}{m^2} \left(v_k^y \right)^2 \left( \vb k \cdot \vb Q\right)  \left( \vb q \cdot \vb Q\right) \left(G_{k,\alpha}^2 G_{k+q,\beta}^4 + 2 G_{k,\alpha}^3 G_{k+q,\beta}^3 +  G_{k,\alpha}^4 G_{k+q,\beta}^2  \right)\\
\tilde G_1^{d} &= \frac{1}{2m^2} \left(v_k^y \right)^2  \left( \vb q \cdot \vb Q\right)^2 \left(G_{k,\alpha}^2 G_{k+q,\beta}^4 + 2 G_{k,\alpha}^3 G_{k+q,\beta}^3 \right).
\end{align}
We assume that $q \ll k_F$ so that $\e_{k} = v_F(k-k_F)$, $\e_{k+q} = \e_k + \vb{v}_F \cdot \vb q$, and $v_k^y = v_F^{y} $. Doing so, we can immediately see that both $\tilde G_1^{a}$ and $\tilde G_1^{b}$ vanish after integration over $\e_k$. In addition, one can show that $\tilde G_1^{c}$ is odd over $\Omega_m$, so it too will vanish. This leaves only $\tilde G_1^{d}$. Now, evaluating $\tilde G_2$ in a similar way, we get
\begin{align}
\tilde G_2^a &= \frac{1}{2m^2} v_k^y v_q^y \left( \vb k \cdot \vb Q \right)^2 \left(G_{k,\alpha}^4 G_{k+q,\beta}^2 + G_{k,\alpha}^2 G_{k+q,\beta}^4 \right)\\
\tilde G_2^b &= \frac{Q^2}{2m} v_k^y v_q^y  \left(G_{k,\alpha}^3 G_{k+q,\beta}^2 + G_{k,\alpha}^2 G_{k+q,\beta}^3 \right)\\
\tilde G_2^c &= \frac{1}{m^2} v_k^y v_q^y \left( \vb k \cdot \vb Q \right) \left( \vb q \cdot \vb Q \right) G_{k,\alpha}^2 G_{k+q,\beta}^4\\
\tilde G_2^d &= \frac{1}{2m^2} v_k^y v_q^y \left( \vb q \cdot \vb Q \right)^2  G_{k,\alpha}^2 G_{k+q,\beta}^4
\end{align}
As before, $\tilde G_2^b$ vanishes after integration over $\e_k$. In addition, both $\tilde G_2^a$ and $\tilde G_2^d$ are odd $\Omega_m$, so they too will not contribute. This leaves solely $\tilde G_2^c$. The total contribution will then come from only $\tilde G_2^d$ and $\tilde G_2^c$. After some simplification by doing integration by parts on the integral over $\e_k$, we find
\begin{align}
\delta \Pi^{JJ}_\perp (H) = \frac{2U^2}{m^2} \sum_{\alpha \neq \beta} \int_{k,q} \Pi^{\alpha \beta}(q,\Omega_m) \left( \vb q \cdot \vb Q \right)v_k^y \left( \left( \vb q \cdot \vb Q \right) v_k^y -  \left( \vb k \cdot \vb Q \right) v_q^y \right)  \left( G_k^{\alpha} \right)^5 G_{k+q}^{\beta}
\end{align}
Noting that, to leading order, the factor of $\vb k$ is simply $k_F \hat k$, then integrating over $\e_k$, and lastly over $\omega_n$ and both angles, we obtain Eq. (\ref{magfieldresult}) in the main text.

\subsection{Verification of the result for  $\delta \chi_{dia} (H)$}

Here we verify Eq. (\ref{magfieldresult}) by computing $\delta \chi_{dia} (H)$ numerically, not restricting to small $q$.
The process involves evaluating the first several integrals and/or sums analytically until we are just left with $q$ and $\Omega_m$. We then evaluate these terms numerically, and compare the result with
(\ref{magfieldresult}). The numerical evaluations were done in Mathematica using the PrincipalValue option to avoid complications from points where $\Omega \rightarrow 0$ and $q \rightarrow 2k_F$, where the integral is singular but convergent in the principal value sense.

 We remind that the transverse current-current correlator can be written as
\begin{align}
\nonumber \Pi_\perp(H)=&U^2 \sum_{\alpha \neq \beta} \int_{k,p}  \Big[\Pi^{\alpha \beta} (q,\Omega_m) \Big( 2 (v_k^y)^2 \left(G_{k+Q/2}^{\alpha} \right)^2 G_{k-Q/2}^{\alpha} G_{k+q+Q/2}^{\beta} \\
& \hspace{140pt}+  v_k^y v_{k+q}^y G_{k+Q/2}^{\alpha} G_{k-Q/2}^{\alpha} G_{k+q+Q/2}^{\beta} G_{k+q-Q/2}^{\beta}\Big)\Big]
\end{align}
We first integrate over $\omega_n$, then series expand to order $Q^2$, and then integrate over angles and over $k$.  Doing this, we obtain
\begin{align}
\nonumber \delta \chi_{dia}(H) = &U^2 N_F^2 \chi_{dia}^0 \int_{0}^{\infty} d \Omega_m \int_{0}^{\infty} d q \frac{1}{8 \pi q^3}\left(\sqrt{1-\alpha ^2-\tilde H}-\sqrt{1-\beta ^2+ \tilde H}\right) \\
\label{eq1} \times \Bigg(&\frac{3 \tilde H^2-2 \tilde H \left(q^2+3 \alpha q+3\right)+\alpha ^2 q^2+2 q^2+6 \alpha  q+3}{\left(1-\alpha ^2- \tilde H\right){}^{5/2}}\\
\nonumber &-\frac{3 \tilde H^2+2 \tilde H \left(q^2+3 \beta q+3\right)+\beta ^2 q^2+2 q^2+6 \beta  q+3}{\left(1-\beta ^2+ \tilde H\right){}^{5/2}}\Bigg) +  (\tilde H\rightarrow - \tilde H),
\end{align}
where  $\tilde H = \mu_B H/E_F$, $\alpha = \frac{i \Omega_m + \tilde H}{q} - \frac{q}{2}$ and $\beta = -\frac{i \Omega_m + \tilde H}{q} - \frac{q}{2}.$ We then subtract off the $H=0$ term to obtain $\delta \chi_{dia} (H)$, and integrate numerically over $q$ and $\Omega_m$ for $\mu_B H$ ranging from $-.02 E_F$ to $.02 E_F$. In this calculation we do not restrict to small $q$.    The result is shown in Fig. \ref{finite_H} of the main text.  It perfectly matches Eq. (\ref{magfieldresult}).

\section{Evaluation of $\delta \chi_{dia} (T)$}
\label{numericT}
Here we show the procedure for evaluating the non-analytic term in the diamagnetic susceptibility at a finite $T$ and $H=0$. We first do summation over internal fermionic frequencies and then expand out to order $Q^2$, as in the case of $T=0$. Once the series expansion is done, the remaining integrals over angles are elementary. The resulting expression takes an unwieldy form, consisting of the sum of terms proportional to the Fermi distribution function and its derivatives. However, it can be simplified to something more manageable by doing
the subsequent integration over fermionic momenta by parts.
Reducing the dependence of the fermionic distribution to Fermi functions, we obtain
\begin{align}
\delta \chi_{dia} (T)= \frac{3}{2} U^2 N_F^2 \chi_{dia}^0 m T \sum_{n=-\infty}^{\infty} \int_{0}^{\infty} dq
 I_p (q) I_k (q)
\label{ss_15_a}
\end{align}
where
\begin{align}
&I_p = \int_{0}^{\infty} d p p\left(-\frac{n_F\left(\e_p\right)}{\sqrt{\alpha ^2-p^2}}-\frac{n_F\left(\e_p\right)}{ \sqrt{\left( \alpha ^*\right) ^2-p^2}} \right) \\
&I_k =
\int_{0}^{\infty} d k k \left(\frac{n_F\left(\e_k\right) \left(\alpha ^2 \left(4 k^2+3 q^2\right)+k^2 \left(k^2+2 q^2\right)+6 \alpha  q k^2 +4 \alpha ^3 q\right)}{q^3 \left(\alpha ^2-k^2\right){}^{7/2}} +c.c.\right),
\label{ss_15}
\end{align}
and $\alpha = \frac{i m \Omega_n}{q} - \frac{q}{2}$.
To proceed, we examine separately the contributions from small bosonic $q$ and from $|q| \approx 2k_F$. For the $q=0$ contribution we show that the non-analytic $O(T)$ term comes from the difference between summation and integration over bosonic Matsubara frequencies, while fermionic distribution functions can be approximated by step functions.
For the $q=2k_F$ contribution, the behavior of the Fermi functions near the Fermi surface become important, and we do no approximate it as a step function.

 \subsection{Contribution from small $q$.}

We assume and then verify that a nonanalytic contribution to $\chi^{q=0}_{dia}$ comes from
$v_F q \sim \Omega_n \sim T$. i.e., from $\Omega_n/q = \mathcal{O}(1)$.
 integrals over $k$ and $p$.
 Using this, we approximate $\alpha ^2-p^2$ by $-\frac{m^2 \Omega_n^2}{q^2} - p^2 + i m \Omega_n$ in the integral over $p$ and do the same in the integral over $k$.
  The integral over $p$ can be written as
\begin{align}
I_p = \int_{0}^{k_F}
 dp p \,   \left(-\frac{1}{\sqrt{-\frac{m^2 \Omega_n^2}{q^2} - p^2 - i m \Omega_n}}-\frac{1}{\sqrt{-\frac{m^2 \Omega_n^2}{q^2} - p^2 + i m \Omega_n}} \right),
\end{align}
Expanding to leading order in $\Omega_n$, we find
\begin{align}
I_p = -\int_{0}^{
k_F}
 dp p \, \frac{m |\Omega_n|}{\left(p^2 + \frac{m^2 \Omega_n^2}{q^2}\right)^{3/2}}
\end{align}
Integrating over $p$, we find
\begin{align}
I_p = q \left(-1 + \frac{|\Omega_n|}{\sqrt{v_F^2 q^2 + \Omega_n^2}} \right).
\end{align}
 We next do similar analysis of the integral over $k$.
Keeping terms of order one and of order $\Omega_n$
 in the numerator, we obtain
\begin{align}
I_k = \int_{0}^{k_F} dk k \left( \frac{k^4 + 2 i m \Omega_n k^2 - 4 k^2 m^2 \Omega_n^2/q^2 - 4 i m^3 \Omega_n^3/q^2}{q^3\left(-\frac{m^2 \Omega_n^2}{q^2} -k^2 - im \Omega_n \right)^{7/2}}+c.c. \right)
\end{align}
Expanding further the denominator to leading order in $\Omega_n$, we find after simple algebra
\begin{align}
I_k = \int_{0}^{k_F}dk k \frac{m|\Omega_n|\left(24 k^2 m^2 \Omega_n^2/q^2 - 3 k^4 - 8 m^4 \Omega_n^4/q^4 \right)}{q^3 \left(k^2 + \frac{m^2\Omega_n^2}{q^2}\right)^{9/2}}
\end{align}
Integrating over $k$, we find
\begin{align}
I_k = \frac{|\Omega_n| v_F^2 q ^2 \left(v_F^2 q^2 - 4\Omega_n^2\right)}{m^2\left(v_F^2 q^2 +\Omega_n^2\right)^{7/2}}
\end{align}
Now, combining this with the results from the $p$ integral, we have
\begin{align}
I_k I_p = \frac{|\Omega_n| v_F^2 q ^3 \left(v_F^2 q^2 - 4\Omega_n^2\right)}{m^2\left(v_F^2 q^2 +\Omega_n^2\right)^{7/2}}   \left(\frac{|\Omega_n|}{\sqrt{v_F^2 q^2 + \Omega_n^2}} -1\right)
\end{align}
The term with $-1$ in the last bracket vanishes after integration over $q$, as one can easily verify. Dropping this term and substituting $I_k I_p$ into (\ref{ss_15}),
we obtain
\begin{align}
\delta \chi^{q=0}_{dia} (T)= \frac{3}{2} U^2 N_F^2 \chi_{dia}^0 T &\sum_{n=-\infty}^{\infty} \int_{0}^{\infty} dq \frac{v_F^2 q^3 \Omega_n^2\left(v_F^2 q^2 -4 \Omega_n^2\right)}{m\left(v_F^2 q^2 + \Omega_n^2\right)^{4}}
\label{ss_16}
\end{align}
 Re-expressing this result in terms of the current-current correlator $\Pi_{JJ}$, we obtain the result that we presented in Eq. ($\ref{smallqT}$) in the main text.

In the main text, we evaluated the frequency sum over $n$ and the integral over $q$ by summing over $\Omega_n$ first, subtracting off the $T=0$ contribution, and then integrating over $q$. For completeness, here we demonstrate that one can also obtain the same result by integrating over $q$ first and then summing over Matsubara frequencies. Since this integral is formally divergent, we must institute an upper cutoff $\Lambda$ on the integral over $q$. In addition, there is ambiguity for the $n=0$ term in the Matsubara sum. For any finite $q$, it is easy to see that the $n=0$ term is zero because of $\Omega^2_n$ in the numerator of (\ref{ss_16}). That said, if we integrate over $q$ first, we can see that the $\Omega^2_n$ cancels out because the q-integration yields $1/\Omega^2_n$.  This last term comes from the lower bound of $q$-integration, i.e., from $q = 0+$. This ambiguity can be resolved by formally instituting a lower cutoff to this term. This lower cutoff will not affect any terms with $n\neq 0$, but will eliminate the $n=0$ contribution. A more physically sound method is to evaluate this integral for a finite system, eliminate the $n=0$ term, and then extend the system size to infinity~\cite{chubukov2003nonanalytic}. Once this is done, we have
\begin{align}
\delta \chi^{q=0}_{dia} (T) &= 3 U^2 N_F^2 \chi_{dia}^0 T\sum_{n=1}^{\infty} \int_{0}^{\Lambda} dq \frac{v_F^2 q^3 \Omega_n^2\left(v_F^2 q^2 -4 \Omega_n^2\right)}{m\left(v_F^2 q^2 + \Omega_n^2\right)^{4}} \\
&= - U^2 N_F^2 \chi_{dia}^0 \frac{T}{4 E_F} \sum_{n=1}^{\infty} \frac{\Lambda^4\left(\Lambda^2 + 6 \Omega_n^2\right)}{\left(\Omega_n^2 + \Lambda^2\right)^3}
\end{align}
The sum over Matsubara frequencies can be evaluated analytically, and gives
\begin{align}
\sum_{n=1}^{\infty} \frac{\Lambda^4\left(\Lambda^2 + 6 \Omega_n^2\right)}{\left(\Omega_n^2 + \Lambda^2\right)^3} = \frac{-64 T^3+36 \Lambda  T^2 \coth \left(\frac{\Lambda }{2 T}\right)-5 \Lambda ^3 \sinh \left(\frac{\Lambda }{T}\right) \text{csch}^4\left(\frac{\Lambda }{2 T}\right)+18 \Lambda ^2 T \text{csch}^2\left(\frac{\Lambda }{2 T}\right)}{128 T^3}
\end{align}
In the limit of $T\rightarrow 0$, the above expression is $\frac{9 \Lambda}{32 T}$. Subtracting this term off, and then taking $\Lambda \rightarrow \infty$, we find for the diamagnetic susceptibility.
\begin{align}
\delta \chi^{q=0}_{dia} (T) &= U^2 N_F^2 \chi_{dia}^0 \frac{T}{8E_F},
\label{eee}
\end{align}
This is the same expression as Eq. (\ref{nnn_1}) in the main text.

\subsection{Contribution from $q \approx 2k_F$.}
\label{appT2kf}
To calculate the $2k_F$ term, we take the above expression for $I_k$, Eq. (\ref{ss_15}) and twice integrate it by parts. We obtain
\begin{align}
I_k = -\int_{0}^{\infty} dk k n_F''(\e_k) \left( \frac{8\alpha^4 + 3 k^4 +4 a^3 q - 6 \alpha k^2 q - 2 k^2q^2 + \alpha^2 (q^2-12k^2)}{3 q^3 m^2 (\alpha^2 -k^2)^{3/2}} +c.c.\right)
\end{align}
Now, we can define variables $x = k^2-k_F^2$, $y=p^2-k_F^2$, and $z = q^2/4-k_F^2$. Lastly, we rescale $x$, $y$, and $z$ by $mT$, and expand in powers of $T$. Then, the total expression is
\begin{align}
\nonumber \delta \chi_{dia}^{2k_F} (T)= \frac{1}{32} U^2 N_F^2 \chi_{dia}^0 T \sum_{n=-\infty}^{\infty} \int_{-\frac{k_F^2}{mT}}^{\infty} dx dy dz n_F(yT/2) n_F''(xT/2) \left(\frac{1}{\sqrt{z - y - i \Omega_n}} + c.c \right) \\ \times \left( - \frac{T}{2\left(z-x-2 \pi i n\right)^{3/2}} + \frac{mT^2}{k_F^2\sqrt{z-x-2 \pi i n}} +\mathcal{O}(T^3) + c.c. \right).
\end{align}
We note that $n_F''$ is of order $T^{-2}$, so to obtain the linear in $T$ team, we must take the second term. Examining this term, we find
\begin{align}
\nonumber \delta \chi_{dia}^{2k_F} (T)=
\frac{1}{64} U^2 N_F^2 \chi_{dia}^0 \frac{T}{E_F} \sum_{n=-\infty}^{\infty} \int_{-\infty}^{\infty} dx dy dz n_F(yT/2) n_F''(xT/2) T^2 \left(\frac{1}{\sqrt{z - y - 2\pi i n}} + c.c \right) \\ \times \left( \frac{1}{\sqrt{z-x-2\pi i n}}+ c.c. \right)
\end{align}
We can then define shift $a=z-x$ and $b=z-y$ to eliminate the $z$ dependence in the square root terms. Then, integrating over $z$ in the two Fermi functions, we find
\begin{align}
\nonumber \delta \chi_{dia}^{2k_F} (T)=
\frac{1}{64} U^2 N_F^2 \chi_{dia}^0 \frac{T}{E_F}\sum_{n=-\infty}^{\infty} \int_{-\infty}^{\infty} da db \;g(a-b) \left(\frac{1}{\sqrt{b - 2\pi i n}} + c.c \right) \left( \frac{1}{\sqrt{a-2\pi i n}}+ c.c. \right),
\end{align}
where we have defined
\begin{align}
g(x) =  \csch^2 \left(\frac{x}{4} \right) \left(1 - \frac{x}{4}\coth \frac{x}{4} \right)
\end{align}
We can then shift $a$ to $a+b$, and integrate over $b$. The integral over $b$ is formally divergent, so we institute a cutoff $\Lambda/T$ at the upper and lower bounds.
 Simultaneously, one has to restrict the summation over $n$ to $|n| < n_{max}$, where $n_{max} = \Lambda/(2\pi T) -1/2$ (Ref.~\cite{PhysRevB.107.144507}).
 Then, we have
\begin{align}
\delta \chi_{dia}^{2k_F} (T)=
\frac{1}{32} U^2 N_F^2 \chi_{dia}^0 \frac{T}{E_F} \sum_{n=-n_{max}}^{n_{max}} S(n),
\end{align}
where
\begin{align}
S(n) = \int_{-\infty}^{\infty} dxg(x) \left( 2 \log \Lambda/T - \log \left( \pi^2 n^2 +\frac{x^2}{16} \right)\right)
\end{align}
We evaluate the $n=0$ and the $n\neq0$ terms separately. For the $n=0$ term, we have
\begin{align}
S(0) = & \int_{-\infty}^{\infty} dx g(x) \left(2\log \Lambda/T - \log \left(\frac{x^2}{16} \right) \right) \\
= & \left( -8 \log \Lambda/T -4 \int_{0}^{\infty} dx g(x) \log (x/4) \right)
\end{align}
 For the contribution from finite $n$ we have
\begin{align}
\sum_{n\neq 0} S(n) = 2\sum_{n=1}^{n_{max}} \int_{-\infty}^{\infty} dx g(x) \left( 2 \log \Lambda/T - \log \left( \pi^2 n^2 +\frac{x^2}{16} \right) \right) \\
= 2 \sum_{n=1}^{n_{max}} \int_{-\infty}^{\infty} dx g(x) \left( 2 \log \Lambda/(\pi T) - 2\log n - \log \left( 1 +\frac{x^2}{16\pi^2n^2} \right) \right)
\label{bbb}
\end{align}
Using $\int_{-\infty}^{\infty} dx g(x) =-4$ we re-write (\ref{bbb}) as
\begin{align}
\sum_{n\neq 0} S(n)= -2\sum_{n=1}^{n_{max}}\left( 8\log \Lambda/(\pi T) - 8 \log n + \int_{-\infty}^{\infty} dx g(x) \log \left( 1 +\frac{x^2}{16\pi^2n^2} \right) \right)
\end{align}
 Using
\beq
\sum_{1}^{n_{max}} \log{n} = (n_{max} +1/2) \log{\left((n_{max} +1/2)/e\right)} + \frac{1}{2} \log{2\pi}
\eeq
we find
\begin{align}
\sum_{n\neq 0} S(n)=  8 \log \left(\frac{2 \Lambda}{T} \right) - \sum_{n=1}^{n_{max}} \int_{-\infty}^{\infty} dx g(x) \log \left(1+ \frac{x^2}{(4\pi n)^2}\right)
\end{align}
Adding together the $n=0$ and $n\neq 0$ terms, we find
\begin{align}
\delta \chi_{dia}^{2k_F} (T)= -
\frac{1}{8} U^2 N_F^2 \chi_{dia}^0 \frac{T}{E_F}
 \left(
\int_{0}^{\infty} g(x) \log (x/2) + \sum_{n=1}^{n_{max}} \int_{0}^{\infty} dx g(x) \log \left( 1 + \frac{x^2}{\left( 4 \pi n\right)^2} \right) \right)
\label{ccc}
\end{align}
The summation over $n$ can be now safely extended to infinity.
Next,
\begin{align}
 \sum_{n=1}^{\infty}
 \log \left( 1 + \frac{x^2}{\left( 4 \pi n\right)^2} \right) =
 \sum_{m=1}^\infty \frac{\zeta (2m) (x/(4\pi))^{2m} (-1)^{m+1}}{m} = \log{\frac{\sinh(x/4)}{(x/4)}}
 \label{ddd}
 \end{align}
where $\zeta (2m)$ is the Riemann Zeta function of an even integer argument.  The last line in (\ref{ddd}) can be verified by expanding $\log{\frac{\sinh(x/4)}{(x/4)}}$ in powers of $x^2$ and comparing terms order by order.

Substituting (\ref{ddd}) into (\ref{ccc}), we obtain
\begin{align}
\delta \chi_{dia}^{2k_F} (T)= -
\frac{1}{8} U^2 N_F^2 \chi_{dia}^0 \frac{T}{E_F}
\int_{0}^{\infty} g(x) \log{(2\sinh(x/4))}
\label{ccc_1}
\end{align}
The product $g(x) \log{(2\sinh(x/4))}$ can be re-expressed as a total derivative, hence the integral can be evaluated exactly.  The result is $\int_{0}^{\infty} g(x) \log{(2\sinh(x/4))} =-1$. The final result is then
\begin{align}
\delta \chi_{dia}^{2k_F} (T)= \frac{1}{8} U^2 N_F^2 \chi_{dia}^0 \frac{T}{E_F}.
\end{align}
Comparing with (\ref{eee}), we see that the $q=2k_F$ contribution to the linear in $T$ term in $\chi_{dia}$ is equal to the contribution from small $q$.
\twocolumngrid
\bibliography{combinedbibliography}
\end{document}